\documentclass[conference]{IEEEtran}
\IEEEoverridecommandlockouts
\usepackage{cite}
\usepackage{amsmath,amssymb,amsfonts}
\usepackage{algorithmic}
\usepackage{graphicx}
\usepackage{textcomp}
\usepackage{xcolor}

\usepackage{fancyhdr}
\usepackage{amsmath,amssymb,amsfonts}
\usepackage{algorithmic}
\usepackage{algorithmic}
\usepackage{graphicx}
\usepackage{textcomp}
\usepackage{xcolor}
\usepackage{svg}
\usepackage{multirow}
\usepackage{indentfirst}
\usepackage{array}
\usepackage{tabularray}
\usepackage{pgfplots}
\usepackage[linesnumbered]{algorithm2e}
\usepackage{calc}
\usepackage{subcaption}
\usepackage{enumitem}
\usepackage{setspace}
\usepackage{pgfplots}
\usepackage{makecell}
\usepackage{booktabs} % for better looking tables
\usepackage{array}
\RestyleAlgo{ruled}
\usepackage{amssymb} % for the \checkmark symbol
\usepackage{pifont}  % for the \ding command
\usepackage{marvosym} % for mail symbol 
\usepackage{pgfplots}
\usepackage{subcaption} % 用于并排放置多个图形
\usepackage{graphicx} % 可能需要用于调整大小
\usepackage{CJKutf8}
\usepackage{adjustbox}
\usepackage{subfig}
\usepackage{footnote}

\SetKwInOut{KwIn}{Input}
\SetKwInOut{KwOut}{Output}
\usepackage[a4paper, total={184mm,239mm}]{geometry}

\def\BibTeX{{\rm B\kern-.05em{\sc i\kern-.025em b}\kern-.08em
    T\kern-.1667em\lower.7ex\hbox{E}\kern-.125emX}}
\begin{document}

\title{

RIROS: A Parallel \underline{R}TL Fault S\underline{I}mulation F\underline{R}amework with Tw\underline{O}-Dimensional Parallelism and Unified \underline{S}chedule
% \\
}
\author{\IEEEauthorblockN{
Jiaping Tang\textsuperscript{*}\textsuperscript{\dag}\textsuperscript{\ddag},
Jianan Mu\textsuperscript{*}\textsuperscript{\dag}\textsuperscript{\Letter}, 
Zizhen Liu\textsuperscript{*}\textsuperscript{\dag}, 
Ge Yu\textsuperscript{*}\textsuperscript{\dag}\textsuperscript{\ddag},
Tenghui Hua\textsuperscript{*}\textsuperscript{\dag}\textsuperscript{\ddag},
Bin Sun\textsuperscript{*}\textsuperscript{\dag}\textsuperscript{\ddag},
Silin Liu\textsuperscript{*}\textsuperscript{\dag},
\\
Jing Ye\textsuperscript{*}\textsuperscript{\dag}\textsuperscript{\ddag},
Huawei Li\textsuperscript{*}\textsuperscript{\dag}\textsuperscript{\ddag}\textsuperscript{\Letter}
}
\IEEEauthorblockA{
\textsuperscript{*}\textit{State Key Lab of Processors, Institute of Computing Technology, Chinese Academy of Sciences, Beijing, China} \\
\textsuperscript{\dag}\textit{University of Chinese Academy of Sciences, Beijing, China}\\
\textsuperscript{\ddag}\textit{CASTEST Co., Ltd., Beijing, China}\\
\{tangjiaping22s, mujianan, liuzizhen, yuge23s, huatenghui24s, sunbin23s, liusilin23s, yejing, lihuawei\}@ict.ac.cn}
% \and
% \and
% \and
% \and
% \and
}

\maketitle

\begin{abstract}
With the rapid development of safety-critical applications such as autonomous driving and embodied intelligence, the functional safety of the corresponding electronic chips becomes more critical. Ensuring chip functional safety requires performing a large number of time-consuming RTL fault simulations during the design phase, significantly increasing the verification cycle.
% 为了满足time-to-market的需求同时确保芯片得到充足的验证，并行加速RTL故障仿真很有必要。
To meet time-to-market demands while ensuring thorough chip verification, parallel acceleration of RTL fault simulation is necessary.
% 多核加速的核心在于减少任务之间的同步开销，而目前的RTL仿真方法主要依赖静态划分方法来最小化任务的同步开销同时实现任务之间的负载均衡。
% 然而，我们发现这种静态划分方法无法适用于RTL故障仿真。根本原因在于故障仿真中故障的传播路径具备动态特征，这使得任务的计算负载表现出不确定性，而静态划分需要依赖相对稳定且确定的负载。
% The key to multi-core acceleration is reducing synchronization overhead between tasks. Current RTL simulation methods rely on static partitioning to minimize synchronization overhead and achieve load balancing. However, static partitioning is ineffective for RTL fault simulation due to the dynamic nature of fault propagation paths, which cause uncertain task loads, while static partitioning requires relatively stable and deterministic loads.
% 然而，由于故障传播路径的动态性且不同故障展具备不同的故障传播能力，这使得RTL故障仿真中的任务负载呈现出极大的不均衡性，使得已有的从单一维度，如结构级，并行的方法在RTL故障仿真不适用。
Due to the dynamic nature of fault propagation paths and varying fault propagation capabilities, task loads in RTL fault simulation are highly imbalanced, making traditional single-dimension parallel methods, such as structural-level parallelism, ineffective.
Through an analysis of fault propagation paths and task loads, we identify two types of tasks in RTL fault simulation: tasks that are few in number but high in load, and tasks that are numerous but low in load. Based on this insight, we propose a two-dimensional parallel approach that combines structural-level and fault-level parallelism to minimize bubbles in RTL fault simulation. 
Structural-level parallelism combining with work-stealing mechanism is used to handle the numerous low-load tasks, while fault-level parallelism is applied to split the high-load tasks. 
Besides, we deviate from the traditional serial execution model of computation and global synchronization in RTL simulation by proposing a unified computation/global synchronization scheduling approach, which further eliminates bubbles.
% 通过对传播路径和任务负载的分析，我们发现RTL故障仿真中存在两类任务：数量少但负载高的任务和数量多但负载低的任务。基于这个发现，我们提出了一个结构级和故障级的二维并行方法去最小化RTL故障仿真中的bubble，使用结构并行来处理数量多但负载低的任务；使用故障级并行来对高负载任务做切分。进一步，我们打破了传统的RTL仿真中计算/全局同步串行的执行方式，使用计算/全局同步统一调度来进一步优化消除bubble。
% To address this challenge, we propose an approach that incorporates dynamic partitioning techniques to effectively adapt to the variability in computational loads during fault simulation. By leveraging real-time analysis of task execution and fault propagation, our method dynamically adjusts the distribution of task loads.  
% 另外，我们通过将全局不同细粒度化局部同步来降低RTL故障仿真中的全局同步开销。根据我们的测量，尽管在RTL仿真中它能够被忽略，但是在RTL故障仿真中该全局同步开销不能被忽略。
% Additionally, we reduce the global synchronization overhead in RTL fault simulation by replacing global synchronization with fine-grained local synchronization. Our measurements show that, while this overhead can be ignored in RTL simulation, it cannot be overlooked in RTL fault simulation.
% 最后，我们实现了一个并行RTL故障仿真框架，RIROS。实验结果表明我们能够获得2.6倍的性能提升相比于静态划分方法同时能够减少56%的全局同步开销。
Finally, we implemented a parallel RTL fault simulation framework, RIROS. 
Experimental results show a performance improvement of 7.0$\times$ and 11.0$\times$ compared to the state-of-the-art RTL fault simulation and a commercial tool.

\end{abstract}

\begin{IEEEkeywords}
Parallel RTL fault simulation, Functional safety verification, Fault injection
\end{IEEEkeywords}

\section{Introduction}

\begin{figure}[htbp]
    \centering
    % \setlength{\abovecaptionskip}{-0.00cm}
    % \setlength{\belowcaptionskip}{-0.00cm}
    % \includesvg[inkscapelatex=false, width=\linewidth]{figure/intro15.svg}
    \includegraphics[width=\linewidth]{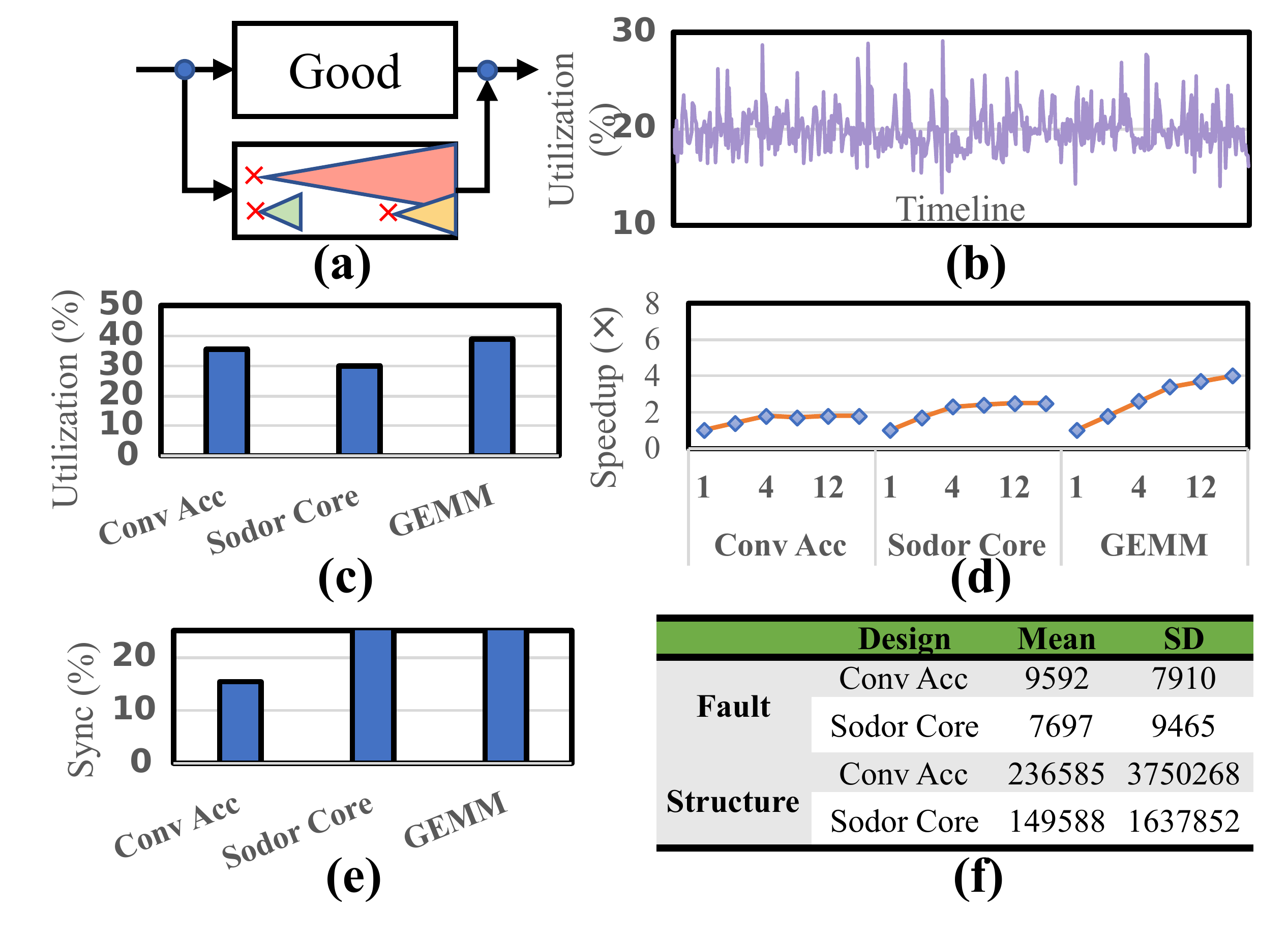}
    \caption{
        (a) Fault simulation diagram with different fault propagation capabilities indicated by the size of cone.
        (b) The CPU utilization for RTL fault simulation exhibits dynamics over time.
        (c) The overall CPU utilization for RTL fault simulation is relatively low.
        (d) The RTL fault simulation performance is poor.
        (e) The overhead of global synchronization in RTL fault simulation cannot be ignored.
        (f) The mean and standard deviation(SD) of task load across different parallelism dimension indicate a significant load imbalance.
    }
    \label{intro_fig}
\end{figure}

% %%%%%intro草稿%%%%%
% 1. RTL级故障仿真，具有重要价值。它是模拟在目标电路上不同潜在故障的传播效应的过程。ISO标准要求汽车芯片等在设计阶段必须覆盖足够数目的潜在故障。
With the rapid advancement of safety-critical applications such as autonomous driving and embodied intelligence, the functional safety of electronic chips has become urgent.
% a significant research focus in both industry and academia. 
The ISO 26262 standard~\cite{ISO} defines specific functional safety requirements for automotive-grade chips.
% , such as microcontroller unit (MCU), application-specific integrated circuit (ASIC), and system-on-chip (SoC).
According to ISO 26262, ensuring functional safety requires extensive fault simulations, which is time-consuming~\cite{long_em, long_2}.
% \mjn{add time data to show it is heavy.}
As shown in Figure~\ref{intro_fig}(a), in contrast to RTL simulation that only requires applying the circuit’s good input value, fault simulation inserts a large number of faults~(indicated by red $\times$) into the circuit. 
These inserted faults are then propagated through the circuit~(as shown by the corresponding cones of red $\times$) to evaluate their impact and obtain the simulation results.

% 2. 看上去故障仿真是一个充满并行性的过程。对于只仿真good输入的RTL逻辑仿真，将电路节点切片并行已取得线性的加速比。但出人意料的是，并行化RTL故障仿真并不容易，如图1.b所示，将已有故障仿真算法运行在16核架构上只能获得3倍左右的加速比，期间CPU核的利用率不到40\%。
% 导致利用率低下的主要原因，在于故障传播效应导致的极大的负载不均衡性。如图1.a所示，在故障仿真中，不同故障的传播能力差异极大，有的故障在传播n层结点之后，就不再对后续结果产生影响，有的故障则可以传播很远。如果不加分辨地将所有故障传播到输出端，则导致整个过程产生高达80\%的冗余执行。故障仿真需要检测在不同电路节点上故障模拟的冗余并进行任务的动态剪枝（这个剪枝需要根据任务执行的结果进行判断，无法在编译层解决），但在削减80\%的冗余执行的同时，这也导致图1.c中展示的故障仿真任务的负载具有动态的不均衡性，最终使得多核执行的CPU平均利用率低下。同时，如图1.d所示，由于减少了仿真任务的执行，电路不同cycle间的同步写回操作的开销占比也得到了提升。

% At first glance, fault simulation seems inherently parallel. In RTL logic simulation of only the good inputs, parallelizing by slicing circuit nodes achieves near-linear speedups. Surprisingly, though, parallelizing RTL fault simulation is not straightforward. 
At first glance, RTL fault simulation appears inherently parallel. 
Although partitioning circuit across cores achieves a near-linear speedup in RTL simulation~\cite{repcut}, extending this parallelism to fault simulation is far from straightforward due to its complex execution dynamics and workload imbalance.
As shown in Figures~\ref{intro_fig}(c)~and~(d), simply running existing fault simulation algorithms on a 16-core platform achieves only about a 4$\times$ speedup, with CPU core utilization below 40\%. 
The primary reason for this low utilization lies in the enormous load imbalance caused by irregular fault propagation behavior. 
As illustrated in Figure~\ref{intro_fig}(a), different faults exhibit varying propagation capability; some faults stop affecting subsequent results after propagating through \textit{n} layers of nodes, while others may propagate far more extensively. 
Indiscriminately propagating all faults to the output results in up to 80\% redundant computations~\cite{ERASER}. 
Consequently, fault simulation must detect and prune redundant simulation at different circuit nodes based on runtime outcomes, while such decisions cannot be made at the compilation stage.
% (a determination that cannot be handled at the compilation layer).
Although pruning eliminates 80\% of redundant executions, it introduces dynamic load imbalance in fault simulation tasks, which leads to low average CPU utilization across multiple cores~(Figure~\ref{intro_fig}(b)). 
Meanwhile, as Figure~\ref{intro_fig}(e) shows, because the number of simulation tasks is reduced, the overhead proportion of global synchronization between cycles has increased significantly.

% 3. 然而，优化RTL故障仿真中的负载动态不平衡，并不是一个容易的问题。如图1.e所示，不管是从故障维度还是电路结构维度进行拆分，RTL故障仿真都面对着任务间的负载不均衡问题：对于故障维度拆分，不同故障的负载差异已然极大;而这也同时导致不同节点间故障执行的负载差异极大（如图1所示）。由于任务间负载差异过大，基于偷线程的动态调度也难以获得线程间的均衡性。因此，已有的RTL仿真中使用的单一维度的任务拆分及管理策略无法有效解决RTL故障仿真的动态负载不均问题。同时，仿真执行与同步操作的串行执行流程中，同步开销占比高而仿真执行阶段利用率低下，整个过程利用率开销极不平衡。

Solving the dynamic load imbalance in RTL fault simulation is far from trivial. 
As Figure~\ref{intro_fig}(f) indicates, whether tasks are decomposed by faults or by circuit structure, RTL fault simulation invariably faces serious load imbalances across different tasks. 
Splitting from the fault dimension alone reveals huge load discrepancies from one fault to another, and likewise, different structural partitions can carry vastly different simulation loads. 
% Because the load variance among tasks is so large, dynamic work-stealing mechanism only supporting task-level stealing struggle to balance workload between cores effectively. 
The substantial load imbalance between tasks makes it difficult for task-level dynamic work-stealing mechanisms to achieve effective load balancing across cores.
Consequently, the single-dimensional task partitioning and management strategies generally employed in RTL simulation are insufficient to address the dynamic load imbalance problem in RTL fault simulation. 
Moreover, in the serial process of simulation execution and global synchronization, high synchronization costs and the low utilization of the execution phase make the overall process highly unbalanced in terms of resource usage.

% 4. To this end，我们提出xx，一个基于多维度拆分和动态调度的RTL故障并行仿真框架。我们结合对于故障传播效应的分析，对于仿真任务，设计了传播效应感知的结构级和故障级的双维度动态切分，从而最小化并行核间负载不均导致的bubble。进一步的，我们提出了多维度的动态管理，我们将原本仿真-同步的操作间串行过程纳入并行化的统一管理，并通过对于细粒度仿真任务的偷线程，实现了更充分的bubble填充。实验结果显示：xxx。

To this end, we propose RIROS, a parallel RTL fault simulation framework that employs multi-dimensional partitioning with unified scheduling. 
Building on an analysis of fault propagation, we introduce a dual-dimensional approach, combining structural-level and fault-level splitting, to minimize bubbles that arise from load imbalances. 
Furthermore, we integrate the previously serial processes of simulation and global synchronization into a unified parallel scheme. By leveraging fine-grained task stealing for global synchronization, we can fill bubbles more efficiently.
Experimental results demonstrate that a 7.0$\times$ and 11.0$\times$ performance improvement compared to the state-of-the art RTL fault simulation and a commercial tool.
The main contributions are summarized as follows.
\begin{itemize}
    \item We identify two types of tasks in RTL fault simulation: high-load tasks and low-load tasks and propose a two-dimensional parallel approach combining structural-level and fault-level parallelism to minimize bubbles.
    \item We identify inherent parallelism opportunities in both the computation and global synchronization stages and propose a unified computation/global synchronization scheduling approach to further reduce bubbles.
    \item We implemented a parallel RTL fault simulation framework using TaskFlow. To the best of our knowledge, this is the first parallel framework for RTL fault simulation.
    \item Comparative experiments demonstrate that our framework achieves a performance improvement of 7.0$\times$ and 11.0$\times$ compared to the state-of-the-art RTL fault simulation and a commercial tool using 16 cores. 
\end{itemize}

\section{Background}
\subsection{\textbf{Parallel Programming}}
% 并行编程模型主要包括loop-based和task-based两类。Loop-based模型通过对循环迭代进行分配以实现数据并行（如OpenMP中的parallel for指令），适用于负载均衡的计算任务。相较之下，task-based模型将程序划分为多个任务，并根据任务间的依赖关系有序执行，适合处理不规则任务的场景。在RTL故障仿真中，故障传播路径具有不确定性，导致计算节点负载不均衡，传统的loop-based并行方法难以充分利用计算资源。因此，本文采用task-based编程模型，通过动态任务划分和依赖管理来有效应对负载不均衡的问题，实现对RTL故障仿真的并行加速。
% 记得添加引用
Parallel programming models are generally classified into two categories: loop-based and task-based. Loop-based models achieve data parallelism by distributing loop iterations (e.g., the \textit{parallel for} directive in OpenMP~\cite{OpenMP}), making them ideal for computational tasks with balanced workloads. In contrast, task-based models divide programs into multiple tasks and execute them in an order determined by their dependencies, which is better suited for handling irregular workloads~\cite{taskflow, async_many_task}. 
% In RTL fault simulation, the uncertainty of fault propagation paths results in imbalanced computational loads across nodes, limiting the effectiveness of traditional loop-based parallel methods in fully utilizing computational resources. 
% To address this, this paper adopts a task-based programming model that utilizes dynamic task partitioning and dependency management to effectively mitigate load imbalances, enabling parallel acceleration of RTL fault simulation.

\subsection{\textbf{RTL Fault Simulation}}
\begin{figure}[htbp]
    \centering
    % \setlength{\abovecaptionskip}{-0.00cm}
    % \setlength{\belowcaptionskip}{-0.00cm}
    % \includesvg[inkscapelatex=false, width=\linewidth]{figure/rtl_graph2.svg}
    \includegraphics[width=\linewidth]{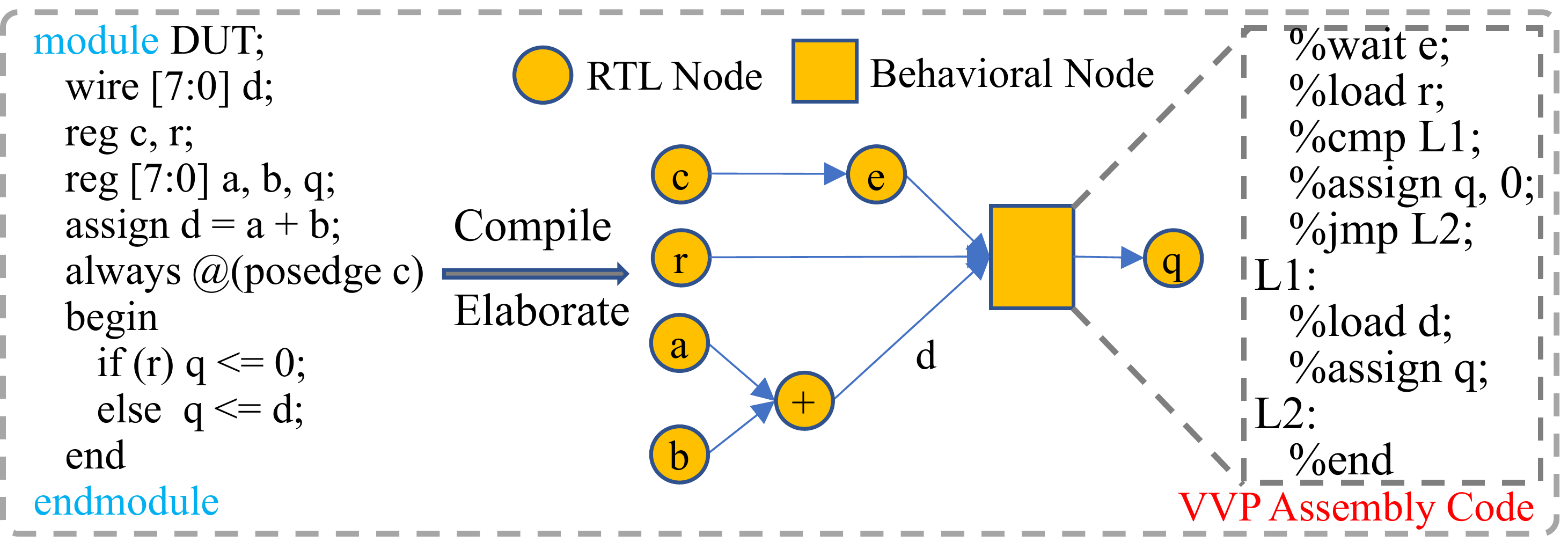}
    \caption{
       An example of RTL code and its corresponding graph representation\cite{iverilog}.
    }
    \label{iverilog_example}
\end{figure}
% RTL代码可以被编译成一个图结构，主要由RTL节点和行为级节点组成。其中，RTL节点用于表示如Reg变量、算术逻辑运算等结构，行为级节点用于表示行为级语句并通过VVP汇编代码来实现其功能。
% RTL node主要由变量（如reg)、算术逻辑运算组成，行为级node表示行为级语句块（如always块）。
% 在RTL仿真中，行为级语句中的非阻塞赋值会在当前仿真结束后统一同步到非阻塞赋值的左部变量上，这些同步后的值会用于下一次迭代仿真。[记得添加Verilog标准]
As shown in Figure \ref{iverilog_example}, 
% RTL code can be compiled into a graph consisting mainly of RTL nodes and behavioral nodes\cite{iverilog}. 
the RTL code can be compiled into a graph consisting mainly of RTL nodes and behavioral nodes. 
RTL nodes represent structures such as variables and arithmetic or logic operations, while behavioral nodes correspond to behavioral statements and are implemented using VVP assembly code~\cite{iverilog}.
% In RTL simulation, non-blocking assignments in behavioral statements are synchronized to their corresponding variables only after the current cycle completes, and these synchronized values are used in the next simulation cycle\cite{verilog_std}. 
In RTL simulation, non-blocking assignments in behavioral statements are synchronized to their left-hand side variables at the end of the current simulation cycle, which is called \textit{global synchronization}, and these synchronized values are used in the next simulation cycle~\cite{verilog_std}. 

% 加一段对故障仿真的解释
% RTL故障仿真指当Wire, Reg或Port上存在故障（如stuck-at-0, stuck-at-1故障）.
RTL fault simulation involves injecting one fault, such as a stuck-at-0 (SA0) fault~\cite{fault_mode}, at wires, registers, or ports. 
% 故障仿真首先需要注入一个故障，然后仿真得到故障存在时，RTL nodes上每个节点的故障值。通过对比无故障时每个节点的正确值能够判断故障能够被检测到。
Then running the simulation to obtain the faulty values for RTL nodes when the fault is present. By comparing these faulty values with the good values obtained in the absence of faults to determine whether the fault can be detected.

The concurrent fault simulation algorithm is widely used in RTL fault simulation~\cite{mozart, multi_cs, ERASER}, which allows simulating multiple faults simultaneously.
 % and adopts an event-driven model and significantly reduces redundant computations. 
In concurrent fault simulation, a node typically includes a good gate and multiple bad gates\footnote{To facilitate understanding, we retain the terms \textit{good gate} and \textit{bad gate} from the concurrent fault simulation algorithm in the context of RTL fault simulation.}~\cite{EDA_book}, with each bad gate induced by a mismatch between the good and faulty values. Both the good gate and bad gates together form its primary computational load for each node. 
However, the computational load at each node in the RTL graph is highly dynamic, posing significant challenges to accelerating RTL fault simulation in parallel environments.

% 在RTL故障仿真中，concurrent故障仿真算法是常用方法。该算法能够同时仿真多个故障。Concurrent故障仿真采用事件驱动模型，能够显著减少冗余计算。
% 然而，算法运行过程中RTL图中每个节点的计算负载动态性较强，导致节点的负载波动较大，而并行加速通常偏好静态负载，使得在并行环境下对RTL故障仿真进行加速充满挑战。
\begin{figure*}[htbp]
    \centering
    % \setlength{\abovecaptionskip}{-0.00cm}
    % \setlength{\belowcaptionskip}{-0.00cm}
    % \includesvg[inkscapelatex=false, width=0.85\linewidth]{figure/motivation0801.svg}
    \includegraphics[ width=0.85\linewidth]{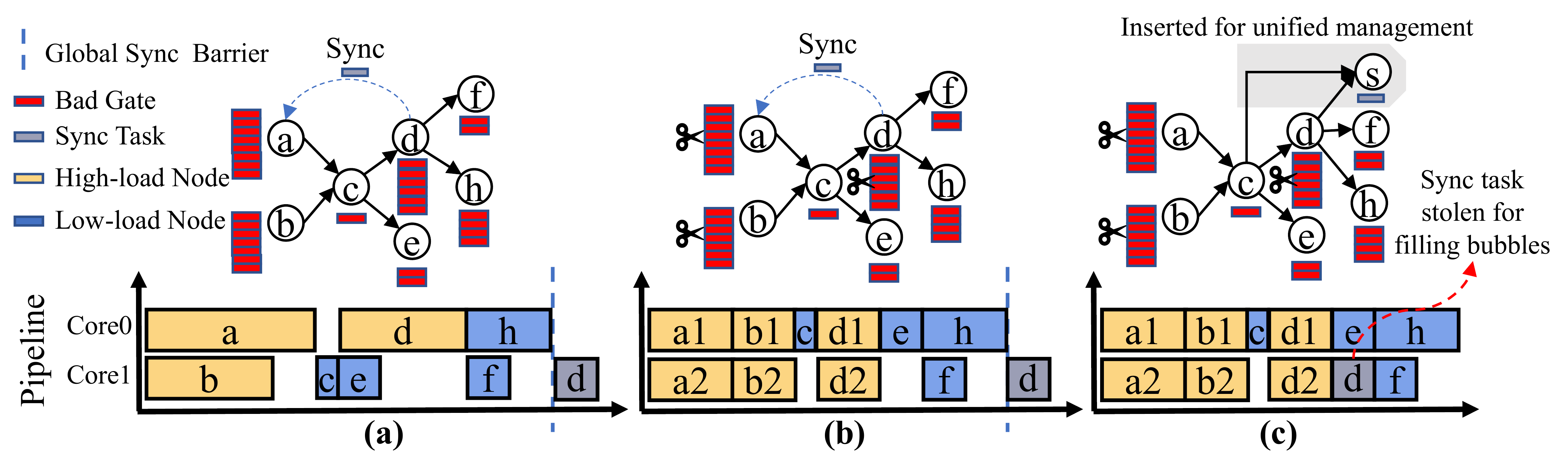}
    \caption{
       Comparison of different parallel dimensions.
       (a): Structural dimension parallelism tends to introduce numerous bubbles;
       (b): Dual-dimensional parallelism of structural dimension and high-load nodes fault dimension reduces bubbles through load balancing;
       (c): Unified task management enables parallel execution of computation and global synchronization, filling bubbles through cross-phase task stealing.
    }
    \label{motivation_fig2}
\end{figure*}
\subsection{\textbf{Related Works}}
% 讲述目前的RTL逻辑仿真的并行加速和RTL故障仿真的单核加速
% 由于该论文是第一篇聚焦于并行加速RTL故障仿真，而在RTL抽象层次已有并行加速RTL仿真和单核RTL故障仿真的工作，所以接下来我们简要描述并行加速RTL仿真和RTL故障仿真相关的工作。
To the best of our knowledge, this paper is the first work to focus on multi-core acceleration of RTL fault simulation. 
So, we provide a brief overview of related work on multi-core RTL simulation and single-core RTL fault simulation.
% 目前多核RTL逻辑仿真采用静态划分算法（verilator, 几篇asplots），静态划分需要能够比较准确地评估图中每个节点的计算负载。而对于逻辑仿真，因为大都采用oblivious仿真算法，所以每个节点的计算负载在编译时能够通过节点执行的指令来较为准确地评估。但是，oblivious仿真算法并不适用于故障仿真，因为这会带来严重的冗余计算。目前的门级故障仿真采用数据并行方式将故障划分给不同的线程并行执行，每个线程每次只处理单个故障。然而，这种以故障维度的数据并行方式并不适用于RTL级故障仿真，因为RTL故障仿真算法天生具备并发故障仿真能力。
% 目前，多核RTL逻辑仿真通常采用静态划分算法（如Verilator及相关ASPDAC、ASPLOS论文中的实现）。静态划分要求在编译时能够准确评估计算图中各节点的计算负载，逻辑仿真中常用的oblivious仿真算法正好满足这一要求，因为它能在编译时根据节点的指令数大致估算出负载。然而，oblivious仿真不适用于故障仿真场景，原因是它会导致大量冗余计算。

Multi-core RTL simulation commonly employs static partitioning algorithms\cite{repcut, manticore,verilator, Parendi}. Static partitioning relies on relatively accurately evaluating the computational load of each node in the graph at compile time. This requirement aligns well with oblivious simulators, which are widely used in RTL simulation, such as Repcut\cite{repcut}, Verilator\cite{verilator}, Parendi\cite{Parendi}. These algorithms can estimate computational loads during compilation based on the number of instructions executed by each node. However, oblivious simulation is unsuitable for fault simulation due to the substantial redundant computations it introduces\cite{essent, ERASER}. 

% 修改RTL代码和使用仿真器内置的命令，such force/release，是常用的RTL故障仿真方法。虽然这些方法可以利用已有的RTL仿真器，但这些方法依赖于大量的手工干预且每次只能仿真单个故障导致性能不足。最近的一些工作集成Concurrent故障方法来支持批故障仿真并消除冗余执行以提高仿真性能。然而，目前已有的RTL故障仿真工作都只支持单核仿真。此外，直接采用多核加速RTL仿真的方法来加速RTL故障仿真并无法取得很好的加速效果。
Modifying RTL code\cite{mefisto ,verify} and using built-in simulator commands\cite{vcs, iverilog}, such as \textit{force/release}, are common approaches for RTL fault simulation. While these methods leverage existing RTL simulators, they rely heavily on manual intervention and can only simulate a single fault at a time, leading to limited performance. 
Recent studies\cite{mozart, multi_cs, ERASER} have integrated concurrent fault methods to support batch fault simulation and eliminate redundant execution, thus enhancing simulation performance. 
However, current RTL fault simulation research is still limited to single-core simulation.
Moreover, directly applying multi-core acceleration techniques from RTL simulation to RTL fault simulation does not yield satisfactory performance gains.
% But for fault simulation, most approaches only support single fault simulation \cite{FI_3 ,mefisto ,verify, HL_1, ISS_1, ISS_2, FSim_1, FSim_2, fast_gp}, leading to suboptimal performance. 

% 去掉门级的故障仿真，有审稿人问了在RTL级为什么要讲门级
% 在门级故障仿真中，当前主流方法采用数据并行方式，将故障静态或动态划分给多个线程，每个线程每次处理单个故障。这种以故障维度进行并行的方式适合门级仿真，但不适用于RTL级故障仿真，因为后者原生具备并发故障仿真的能力，能够在同一仿真迭代中处理多个故障。
% In gate-level fault simulation, the dominant approach utilizes a data-parallel strategy, dividing faults—either statically or dynamically—among multiple threads, with each thread processing one fault at a time\cite{gate_level_fsim1,gate_level_fsim2,gate_level_fsim3}. While this fault-level parallelism works well for gate-level simulation, it is not effective for RTL-level fault simulation. RTL fault simulation inherently supports concurrent fault processing, allowing multiple faults to be simulated simultaneously within the same simulation iteration.

\section{Motivation of RIROS}\label{motivation}
% 观察 - insight - 方法
\begin{figure}[htbp]
    \centering
    \includegraphics[width=\linewidth]{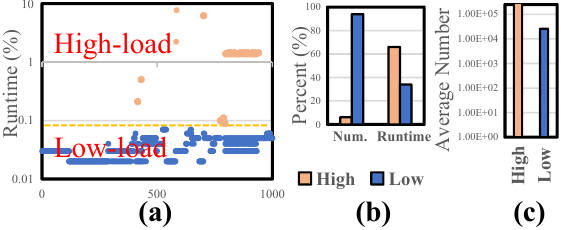}
    % \includesvg[inkscapelatex=false, width=\linewidth]{figure/motivation422_3.svg}
    \caption{
        Performance bottleneck breakdown of Convolution Accelerator~(Conv Acc) design.
       (a): Computation overhead for each node expanded along the topological order~(X-axis);
       % Structural-level computation overhead distribution;
       (b): Distribution of node count and runtime proportion between high-load and low-load nodes;
       % Proportion of high-load and low-load nodes in count and runtime;
       (c): The average number of bad gates in high-load and low-load nodes.
    }
    \label{motivation_fig}
\end{figure}

% 我们对RTL故障传播，不同电路节点上的负载进行了分析。如图2所示，可以明显看到包括2类节点：1类节点数目少，但是其任务负载极高。另一类节点任务负载低，但是这类节点数目很多。
% 我们对这两类节点分别进行评估分析，如图2.b所示，第一类节点，它的节点更大，挂载的故障列表更长。第二类节点，它的节点更小，故障列表相对少，但是在完整电路中，这样的节点数目很多。

% 因此，我们的想法是，结合不同节点的故障负载特性，使用不同的并行维度：
% 如图2.c所示，对于大负载节点，我们按照故障维度进行任务拆分和并行，能够获得较好的并行性。而对于大量的小负载节点，我们按照节点维度进行任务拆分，而由于这些节点的负载相对均衡，结合偷线程的动态调度策略，可以获得较好的均衡效果。这样的多维度并行策略，能够最小化仿真执行的bubble。

% 另一方面，通过图2可以看到，目前的算法使用仿真-同步的串行执行流程。仿真部分由于动态负载不均，多核并行架构具有大量bubble，而同步部分的开销又占比很高。我们对此的想法是，进行不同操作类型间的统一化并行管理机制，并行执行仿真和同步操作，从而最大程度填充整个阶段的bubble。

We analyzed RTL fault propagation and the load distribution across different circuit nodes. As shown in Figures~\ref{motivation_fig}(a) and (b), there are clearly two types of nodes: one type is small in number but carries a very high computational load, while the other type is large in number but has a relatively low per-node load.
% We conducted a separate evaluation of these two node types. 
We further evaluate the two types of node.
As illustrated in Figure~\ref{motivation_fig}(c), the first type includes a large number of bad gates, while the second type features with fewer bad gates.
% , but they are numerous within the overall circuit.

Based on these differing fault-load characteristics, our approach is to adopt different parallelization dimensions. 
Compared to structural-level parallelism shown in Figure~\ref{motivation_fig2}(a), for high-load nodes, we split tasks along the fault dimension to achieve better parallelization depicted in Figure~\ref{motivation_fig2}(b), as there are no dependency constraints between bad gates.
Meanwhile, for the large number of low-load nodes, we partition tasks by node dimension. Because each of these nodes has a relatively similar load, combining with a work-stealing dynamic scheduling strategy can achieve load balance effectively. 
This multi-dimensional parallelization strategy minimizes bubbles during simulation execution.

On the other hand, as shown in Figure~\ref{motivation_fig2}(a) and (b), current algorithms rely on a serial flow of computation followed by synchronization. 
During RTL fault simulation, dynamic load imbalance leads to many bubbles on multi-core architectures, while synchronization overhead remains high. 
We identify fine-grained parallelism opportunities between computation and global synchronization. 
Specifically, once a behavioral task has completed and the synchronized variables are no longer used in the current cycle, synchronization can be safely pre-executed without waiting for the completion of all computations.
Our proposed solution is to introduce a unified parallel management mechanism that executes computations and global synchronization in parallel, as shown in Figure~\ref{motivation_fig2}(c), thereby maximizing the utilization of idle slots.

% \input{motivation_old}

% 按照总分的形式写
\section{RIROS Frameworks}

\begin{figure*}[htbp]
    \centering
    % \setlength{\abovecaptionskip}{-0.00cm}
    % \setlength{\belowcaptionskip}{-0.00cm}
    % \includesvg[inkscapelatex=false, width=0.75\linewidth]{figure/framework4.svg}
    \includegraphics[ width=0.75\linewidth]{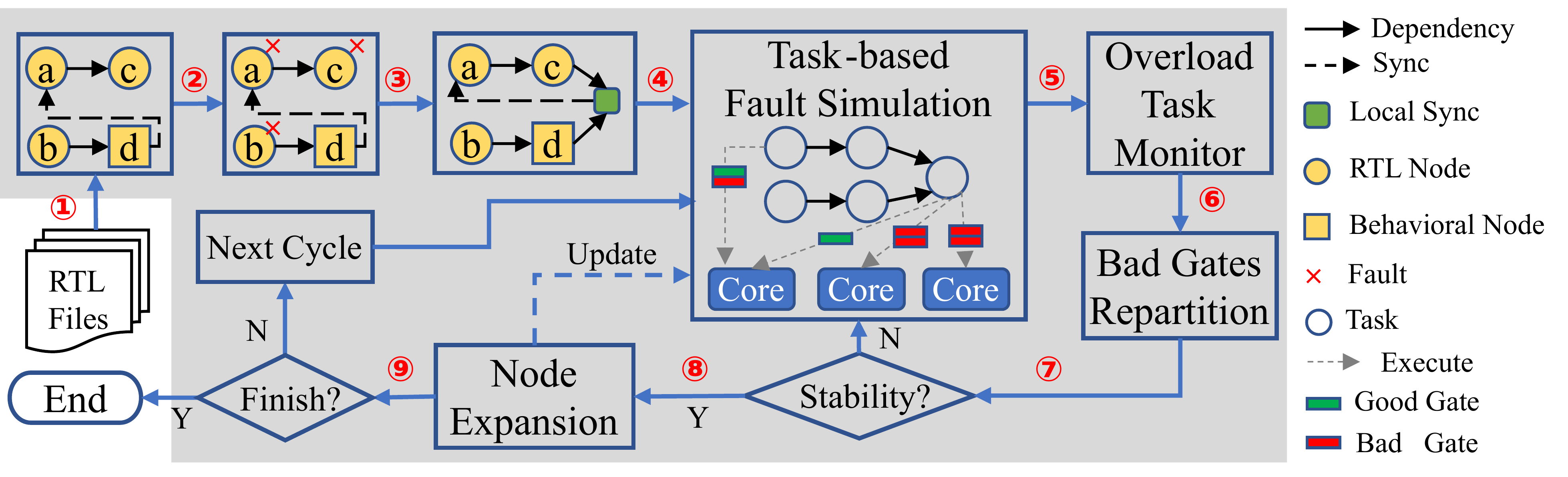}
    \caption{
       Framework of RIROS
    }
    \label{framework}
\end{figure*}

% 如图xxx所示，该框架以RTL文件作为输入，总共包含9个步骤。
% 步骤1将RTL代码compile and elaborate成一个RTL图。
% 步骤2使用统一故障注入方法完成批故障注入。
% 步骤3为行为级代码增加局部同步点，得到一个插入局部同步点的图。
% 步骤4构建Taskflow任务图，并实现Concurrent故障仿真。利用Taskflow执行引擎并行加速故障仿真。
% 步骤5和步骤6通过load-aware task repartitioning方法将高负载节点划分成多个子任务并为子任务分配bad gates，避免RTL故障仿真因高负载节点导致的核空闲问题。
% 步骤7用于判断当前cycle是否仿真结束，如果没有，则进行继续迭代计算图直到电路达到稳态。
% 否则，步骤8扩展高负载节点的任务重分配能力，用于后续仿真的任务重分配。注意，该步骤会更新任务图。
% 步骤9判断仿真是否结束，如果未结束则进行下一cycle的仿真。
% As shown in Fig. \ref{framework}, this framework processes RTL files through eight steps to ensure efficient RTL fault simulation. 
% It begins by compiling and elaborating the RTL code into a computational graph (Step 1), 
% followed by refining the graph with the insertion of local synchronization nodes (Step 2). 
% Concurrent fault simulation is then executed with parallelization to improve performance (Step 3).
% Steps 4 and 5 mitigate high-load node issues through load-aware task repartitioning, breaking them into sub-tasks to reduce thread idling and balance the workload. 
% Step 6 checks whether the current simulation cycle has reached steady state; if not, the computational graph is iteratively processed. 
% Once steady-state is achieved, Step 7 updates the graph to optimize task repartitioning for subsequent cycles. 
% Finally, Step 8 evaluates whether the simulation is complete, advancing to the next cycle if needed. 

As illustrated in Figure \ref{framework}, the framework takes RTL files as input and comprises nine steps. 
Step 1 compiles and elaborates the RTL code to generate a RTL graph. 
Step 2 performs batch fault injection using a unified fault injection method. 
In Step 3, local synchronization nodes are inserted for each behavioral node. 
% Step 4 constructs a task graph using Taskflow\cite{taskflow} and performs concurrent fault simulation, leveraging the Taskflow execution engine for parallel acceleration. 
Step 4 constructs a task graph through a one-to-one mapping from the RTL graph and performs concurrent fault simulation with multi-core. 
Steps 5 and 6 introduce a monitor-based overload-aware task partitioning strategy that splits high-load nodes into multiple sub-tasks and assigns bad gates to each, thereby preventing core idling and balance the workload. 
Step 7 checks whether the simulation has reached a steady state in the current cycle; if not, the task graph continues to iterate. 
Once the circuit stabilizes, Step 8 updates the task graph to optimize task repartitioning for subsequent cycles. 
Finally, Step 9 evaluates whether the simulation is complete, advancing to the next cycle if needed. 
This systematic approach achieves an effective workload balance and substantially improves parallel simulation performance.

\subsection{\textbf{Fault Injection}}\label{fault_injection}
% 如表xxx所示，RIROS使用了统一的故障注入方法用于支持不同故障位置（Wire, Reg和Port）和不同故障类型（SA0, SA1等）的故障注入。对于不同的故障注入位置，我们都统一将其注入到其对应的RTL Node上。具体而言，Wire上的故障注入到该Wire的驱动所对应的RTL Node上；由于Reg存在对应的RTL Node，那么Reg上的故障直接注入到其对应的节点上；对于Port上的故障，会在RTL Graph中先生成一个虚拟节点，然后注入到相应的虚拟节点上。
As shown in Table \ref{faultInjection}, RIROS proposes a unified fault injection approach to support various fault locations (Wire, Reg, and Port) and fault types (SA0, SA1, etc.). 
Regardless of the fault location, all faults are uniformly injected into their corresponding RTL nodes. Specifically, a fault on a wire is injected into the RTL node that drives the wire. Since a reg has a directly associated RTL node, faults on the reg are injected into that node. For port faults, a virtual node is first inserted into the RTL graph, and the fault is then injected into this virtual node.

% 在故障注入后，每个节点都会维护表格用于存储与故障相关的数据。
% fid是一个全局的故障编号；fptr是一个故障指针，null表示故障注入在其他节点上，非null表示故障注入在该节点上。fval存储故障在该节点的故障值。
% 在计算时，首先通过evaluate函数计算故障在节点的故障值，然后再查看节点的故障注入表格查看该故障是否注入到当前节点，如果注入的话，那么调用FaultyVal函数得到最终的故障值f\_out，它是一个虚函数用于支持不同的故障类型。最后，使用f\_out更新faulty machine.
After fault injection, each RTL node maintains a table to store fault-related data. 
\textit{fid} is a global fault identifier. 
\textit{fptr} is a fault pointer; when non-null, it indicates that the fault is injected at the node.
% , while a null value indicates that the fault is injected at others nodes. 
\textit{fval} holds the faulty value at the node.
During fault simulation, for each activated bad gate, the \textit{compute} function is first called to calculate the faulty output value based on faulty input values at the node. The table is then checked using \textit{fid} extracted from the bad gate to determine whether a fault has been injected. If a fault is present, the \textit{faultyVal}, a virtual function designed to support different fault types, is invoked to obtain the final faulty value \textit{f\_out}. Finally, update \textit{fval} with \textit{f\_out}.

\begin{table}
    \centering
    \caption{A unified fault injection method}
    \begin{tabular}{c}
       % \includesvg[inkscapelatex=false, width=0.95\linewidth]{figure/fault_injection4.svg}
       \includegraphics[ width=0.95\linewidth]{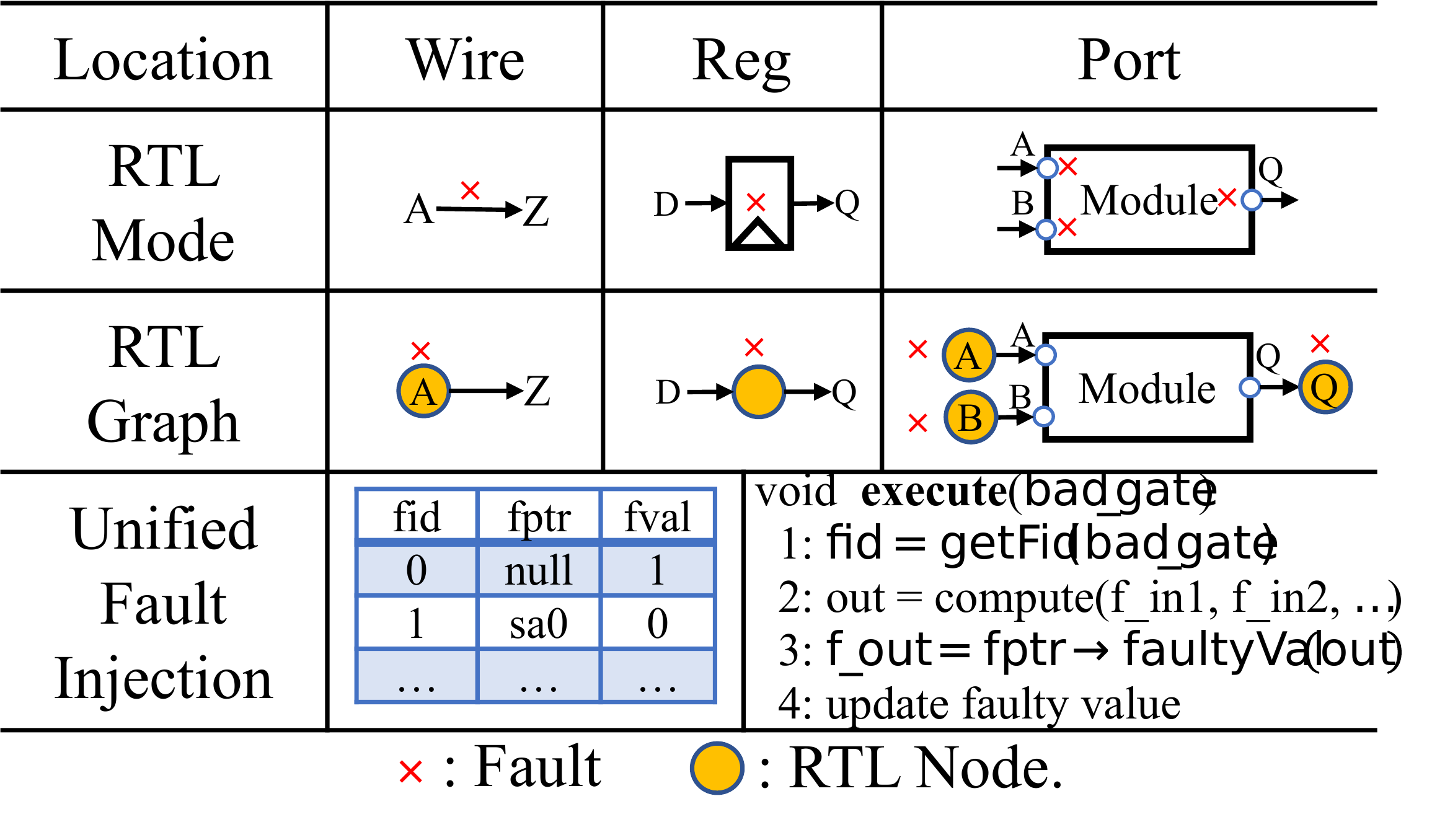}
    \end{tabular}
    % \caption*{}
    \label{faultInjection}
\end{table}

% 框架算法
\subsection{\textbf{Task-based RTL Fault Simulation}}
% 先讲Task-based fault simulation. 
% 在我们框架中，我们舍弃了对事件驱动队列地管理，尽管该队列是事件驱动仿真的核心数据结构。在并行环境下，事件队列作为一个公有数据，对它的写需要涉及大量的同步管理，比如加锁。我们认为这会严重降低并行性能。
In our framework, we eliminate the management of the event-driven queue, despite it being a core data structure in concurrent fault simulation. In parallel environments, the event queue, as a shared resource, requires extensive synchronization mechanisms, such as locking, to handle write operations, which can severely hinder parallel performance. 
% RTL graph到task graph的转换（这个放在框架的描述里）

% 我们使用task-based并行框架来实现RIROS，这里，我们先给出一个总体的概括。
We implement RIROS using a task-based parallel framework. Here, we first provide an overview.
% 算法1展示了我们的故障仿真算法，该算法采用遍历-检查-执行-更新的方式来并行处理计算图中的每个节点。
Algorithm~\ref{alg_execution} presents an overview of the fault simulation algorithm used in RIROS, which parallelizes the processing of each task in the task graph using a \textit{traverse}-\textit{check}-\textit{execute}-\textit{update} workflow.
% 首先我们从任务图中获得一个就绪的任务(line 3)，一个就绪任务指该任务所依赖的所有任务都已执行完。然后，我们会检查该任务依赖的数据是否发生变化，并决定是否需要执行该任务(line 4-5)。如果任务的依赖的数据存在新的值，那么我们根据任务的类型完成具体的执行(line 7-26).最后，将当前任务的输出用于更新该任务的后续任务。
First, we retrieve a ready task from the task graph (line 2), defined as a ready task whose dependencies have been resolved. 
Next, we check whether the task's dependent data have changed to determine if execution is required (lines 3). If new values are detected in the dependencies, the task is executed based on its specific type (lines 4-21), which will be introduced in the following sections.
Finally, the output of task is used to update its successor tasks(line 22).

\begin{algorithm}
\caption{Task-based Fault Simulation} \label{alg_execution}
\KwIn{ TG: A task graph of RTL}
\KwOut{ An expanded task graph }

% overload\_task\_queue $\gets$ an empty queue\;

\While{simulation of current cycle not finish} {

        ready\_task $\gets$ an ready task from TG\;
        % \If{hasn't value change} {
        %     Continue\;
        % }
        check whether the dependence has changed\;
        \Switch{ready\_task}{
            \Case{master task}{
                execute(good\_gate)\;
                distribute bad gates evenly\;
                assign bad gates indices to slaves\;
            }
            \Case{slave task}{
                execute(bad\_gates, \{begin, end\})\;
            }
            \Case{sync task}{
                execute\_sync()\;
            }
            \Else{
                execute(good\_gate)\; 
                execute(bad\_gates)\;
                check whether the task is a high-load task\;
            }
        }
        update successor tasks\;       
}
node expansion for high-load tasks\;
\end{algorithm}

\subsection{\textbf{Two-Dimensional Parallelism}}
\subsubsection{\textbf{Structural Dimension Parallelism}}
% 根据xxx Section所述，RTL故障仿真中存在大量的节点维度的小任务，这意味着1）从节点维度并行可以提供大量的并行机会；2）结合工作窃取机制能够实现不错的负载均衡。
% 在我们的节点级并行中，只要发现节点的依赖节点（前驱节点）已完成计算，那么我们就会将该节点的计算负载(包括good gate和bad gates)加入到任务发射队列中。只要发现存在空闲的核，那么我们就会从发射队列中发射一个节点到核上去执行(算法1中的line 19-24)。
As discussed in Section~\ref{motivation}, there are a large number of low-load tasks at the node level in RTL fault simulation. 
This characteristic makes fine-grained structural-level parallelism particularly effective, as it offers abundant opportunities for concurrent execution. 
% 由于目前的task-based编程天生具备节点级并行能力，所以我们这里只简要描述一下我们的节点级并行方法。然而，需要值的注意的是，我们的节点级并行只作用RTL仿真中的小任务，对于大任务，我们会采用故障级并行方法。
Given that modern task-based programming inherently supports structural-level parallelism, we provide only a concise overview of our structural-level parallel strategy in this section.
% 具体而言，对于所有依赖已经resolved的任务，我们会并行的执行它们。在我们的统一框架中，任务可以是同步任务，也可以是master-slaver任务（对high-load任务采用故障并行所引入），也可以是low-load任务（完成good gate和bad gate的计算）。
% 在任务执行完后，该任务又会触发后继任务的并行执行，直到所有任务都执行完。
Specifically, all tasks whose dependencies have been resolved are executed in parallel. In our unified framework, a task can be a synchronization task~(lines 13-15), a master-slave task~(lines 5-12) introduced by fault dimension parallelism for high-load tasks, or a default task computing the good and bad gates~(lines 16-20).
After a task is completed, it invokes the parallel execution of its successor tasks, continuing this process until all tasks have been executed.
% In our approach, once a node's dependencies (i.e., its predecessor nodes) have completed execution, the node's computation task—covering both good gates and bad gates—is placed into a task dispatch queue. Whenever an idle core is available, a task is immediately dispatched from the queue to the core for execution (lines 19–23 in Algorithm~\ref{alg_execution}).

\subsubsection{\textbf{Fault Dimension Parallelism}}
\begin{figure}[htbp]
    \centering
    \setlength{\abovecaptionskip}{-0.00cm}
    \setlength{\belowcaptionskip}{-0.00cm}
    % \includesvg[inkscapelatex=false, width=\linewidth]{figure/task_partition2.svg}
    \includegraphics[ width=\linewidth]{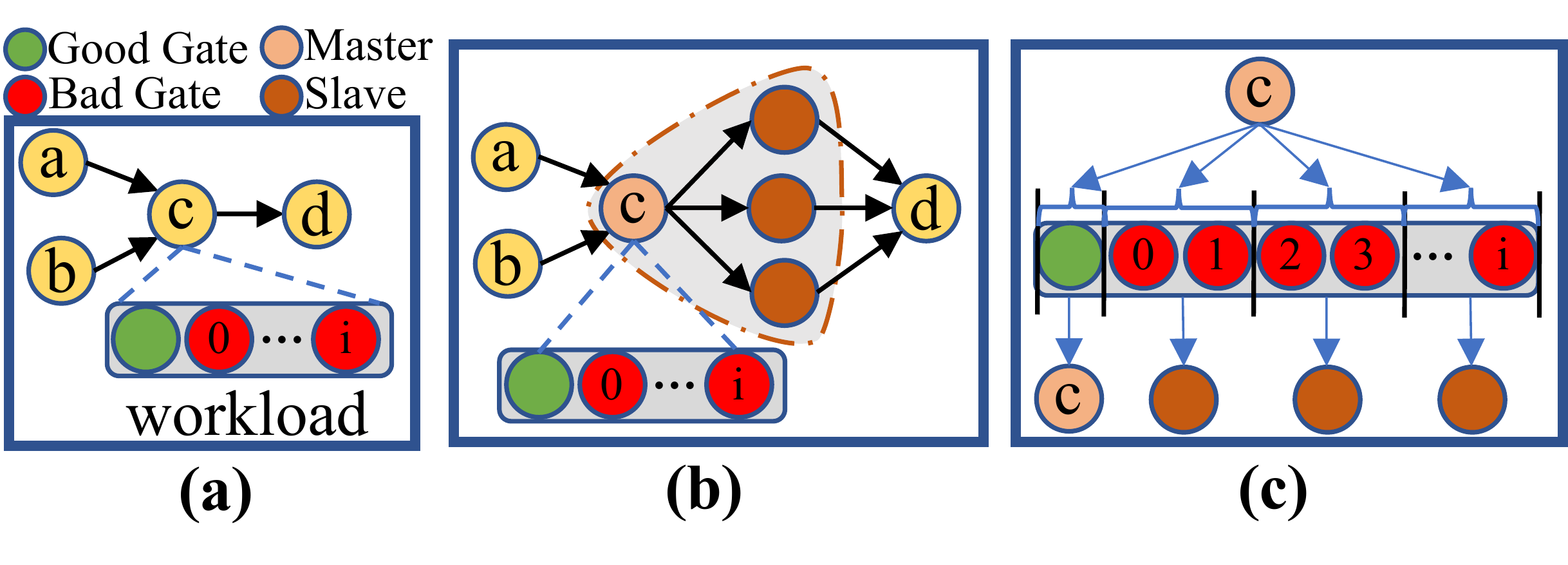}
    \caption{
       An example of fault dimension parallelism.
        (a) Consider a scenario where the task on node \textit{c} becomes overloaded. 
        (b) The overloaded node expands into a master-slave structure, adding new  dependency relationships.  
        (c) The master task distributes the computational load between itself and the slave tasks.
    }
    \label{task_repartition}
\end{figure}
To mitigate the load imbalance caused by heavy-load nodes in structural-level parallelism, we introduce a fault-dimension parallelism approach, which divides overloaded nodes into multiple independent sub-tasks along the fault dimension.
Our task partitioning method adopts a one-master-multi-slave structure~(illustrated in Fig. \ref{task_repartition}(b)) for heavy-load nodes. In this structure, the master task handles the simulation of the good gate, while multiple slave tasks are responsible for simulating the bad gates~(illustrated in Fig. \ref{task_repartition}(c)), effectively distributing the computational workload among slave tasks. 
In concurrent fault simulation, the simulation of a good gate must take precedence to avoid being overwritten by the bad gate simulation. To ensure this constraint, we introduce dependency constraints between the master and slave tasks, ensuring that the master task executes first, followed by the slave tasks. 
Furthermore, the explicit computational load associated with each node in fault simulation allows for precise task partitioning, facilitating load balancing across sub-tasks. 
Specifically, we implement a uniform task allocation strategy within the master task to evenly distribute the computational load of bad gates among the slave tasks. This ensures that the execution load across slave tasks is nearly identical, reducing cores idling and improving parallel performance. 
% Using this fine-grained task partitioning and load balancing mechanism, our method significantly enhances parallel efficiency and minimizes the adverse effects of load imbalance on performance.

% TODO：可以给节点取个名
% 算法xxx。具体来说，我们实时统计每个计算节点的运行时间占比，将超过阈值的节点标记为负载过重节点，并在当前仿真结束后将其扩展为one-master-multi-slave结构，该结构该能够有效地避免动态任务划分带来的成本开销。当一个节点被扩展成one-master-multi-slave节点时，意味着该节点具备负载意识的任务重分配能力，能够增强任务分配的灵活性。注意，这里仅仅只做节点扩展，并没有做具体的任务分配，具体的任务分配会在仿真运行时决定。在运行时，我们从任务图中取出一个就绪执行的节点执行。master节点首先执行good gate任务并统计需要处理的bad gate数量，然后通过轻量级的数组下标索引策略将bad gate任务区间分配给slave节点，避免了动态任务划分的额外开销。具体而言，master节点维护一个任务数组用来表示所有需要被计算的bad gate，然后master节点通过[begin, end)数组下标的形式将任务发送给slave节点。等到slave执行时，每个slave节点根据接收的区间从master中窃取任务并执行任务。
We use a monitor-based approach to identify whether a node qualifies as a heavy-load node. 
Specifically, we dynamically monitor the execution time share of each computational node in real-time. Nodes whose execution time exceeds a predefined threshold are marked as overloaded, and after the current simulation cycle, these heavy-load nodes are expanded into the one-master-multi-slave structure~(line 24). 
% This structure effectively mitigates the overhead associated with dynamic task partitioning.
When a node is expanded into a one-master-multi-slave structure, it gains the capability of fault-dimension parallelism. It is important to note that only the node expansion is performed at this stage; the actual task allocation will be determined during simulation execution.
During simulation, the master node first processes the good gate task and calculates the number of bad gates to be evaluated. Then, using a lightweight array indexing strategy, the master node allocates the bad gate task intervals to the slave nodes~(line 8), avoiding the additional overhead of dynamic task partitioning. Specifically, the master node maintains a task array that lists all the bad gates to be evaluated, and the tasks are sent to the slave nodes using array indices [begin, end). 
% When the slave nodes execute, each one steals a portion of the task based on the received index range and performs tasks.
Each slave node, upon execution, steals a portion of the workload based on its assigned index range and proceeds to perform the corresponding tasks.

% This approach effectively balances the computational load, enhances parallel execution, and minimizes the overhead typically associated with dynamic task division.

\subsection{\textbf{Unified Schedule between Computation and Global Synchronization}}
\begin{figure}[htbp]
    \centering
    \setlength{\abovecaptionskip}{-0.00cm}
    \setlength{\belowcaptionskip}{-0.00cm}
    % \includesvg[inkscapelatex=false, width=0.9\linewidth]{figure/load_sync2.svg}
    \includegraphics[ width=0.9\linewidth]{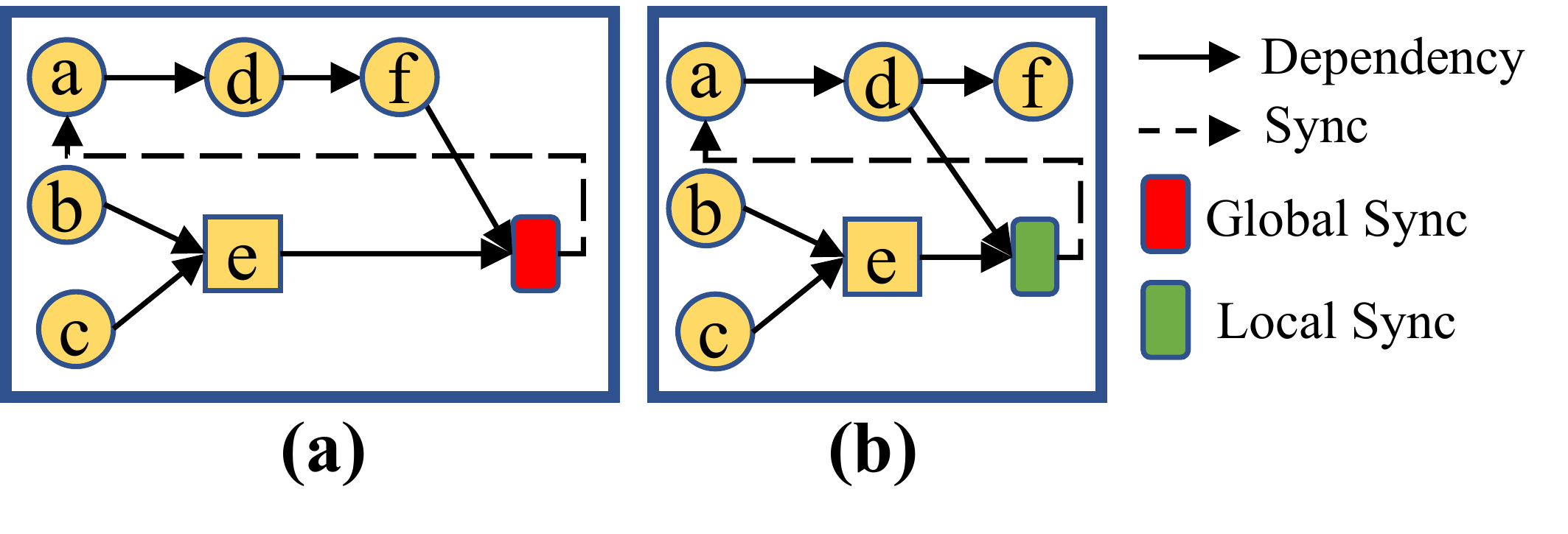}
    \caption{
       Insert fine-grained local synchronization to replace global synchronization.
       (a) Using global synchronization.
       (b) Using local synchronization.
    }
    \label{sync_hidden}
\end{figure}
While structural-level parallelism and fault-dimension parallelism have been successfully implemented, completely eliminating core bubbles during the computation phase remains a significant challenge. Moreover, in RTL fault simulation, the overhead of global synchronization introduces a notable performance bottleneck that cannot be ignored, as illustrated in Figure~\ref{intro_fig}(e).

To optimize the utilization of computation-phase bubbles, we move away from the traditional serial execution model of "computation-global synchronization" and instead integrate both computation and global synchronization into a unified scheduling framework. 
This approach enables the early global synchronization tasks stealing during the computation phase, effectively filling the bubbles. 
However, integrating these two processes within the same scheduling framework is not trivial, as we must ensure the correctness of RTL simulation semantics. Specifically, to avoid simulation errors caused by prematurely using next-cycle values in the current cycle, it is crucial to determine the appropriate timing for local synchronization. Variables can only be synchronized when they have been fully used in the current cycle and will no longer be accessed thereafter. 

To implement a unified schedule, we add dedicated local synchronization tasks for behavioral blocks to support advanced synchronization. We also introduce appropriate dependencies to ensure the correctness of task execution. 
This method seamlessly integrates local synchronization into the task graph, allowing it to be executed automatically according to the task graph's scheduling rules. 
Our process begins by analyzing non-blocking assignment statements in behavioral code blocks to identify synchronization variables. For each synchronization variable, we identify all tasks that are dependent on it. 
These dependent tasks must complete execution before the variable is synchronized; otherwise, the variable might be overwritten with its next-cycle value, causing simulation errors.

\begin{table*}[htbp]
  \centering
  \caption{Performance comparison between ERASER, VCS and RIROS}
  \resizebox{0.95\textwidth}{!}{
  \begin{tabular}{l|rrrr|rrrr|rrr}
    \toprule
    \toprule
    \multirow{3}{*}{\textbf{Benchmark}} & 
    \multicolumn{4}{c|}{\multirow{2}{*}{\textbf{Statistics}}} & 
    \multicolumn{4}{c|}{\multirow{1}{*}{\textbf{Runtime(s)}}} & 
    \multicolumn{3}{c}{\multirow{1}{*}{\textbf{Speed-up}}}
    
    % \multicolumn{3}{c|}{\textbf{Statistics}} &
    % \multicolumn{6}{c|}{\textbf{Runtime}} & 
    % \multicolumn{3}{c}{\textbf{Speed-up}} 
    \\  
    \cline{6-9} & & & & & 
    \multicolumn{1}{c}{ERASER} & 
    \multicolumn{1}{c}{VCS} & 
    \multicolumn{2}{c|}{RIROS} & 
    \multicolumn{3}{c}{RIROS 16 Cores VS.} 
    \\  
     & 
    \multicolumn{1}{c}{\#RTL Nodes} & 
    \multicolumn{1}{c}{\#VVPs} & 
    \multicolumn{1}{c}{\#Stimulus} & 
    \multicolumn{1}{c|}{\#Faults} &
    \multicolumn{1}{c}{1} & 
    \multicolumn{1}{c}{16} & 
    \multicolumn{1}{c}{1} & 
    \multicolumn{1}{c|}{16} & 
    \multicolumn{1}{c}{ERASER} & 
    \multicolumn{1}{c}{VCS} & 
    \multicolumn{1}{c}{1 Core} \\
    \midrule

   \texttt{RISCV Mini} & 1426 & 1306 & 4.0k & 17748  & 
    905.0 s & 3549.6 s & 928.5 s & 198.7 s &
    4.6$\times$ & 17.9$\times$ & 4.7$\times$ \\

    \texttt{Conv Acc} & 1061 & 2140 & 4.0k & 17880  & 
    1188.5 s & 3754.8 s & 1087.0 s & 203.1 s &
    5.9$\times$ & 18.5$\times$ & 5.4$\times$ \\

     \texttt{Sodor Core} & 2842 & 1249 & 4.0k & 4678  & 
    873.4 s & 1075.9 s & 839.6 s & 146.0 s &
    6.0$\times$ & 7.4$\times$ & 5.8$\times$ \\

     \texttt{PicoRV32} & 435 & 3754 & 4.0k & 6898  & 
    449.9 s & 1448.6 s & 301.5 s & 56.4 s &
    8.0$\times$ & 25.7$\times$ & 5.3$\times$ \\
    
    \texttt{GEMM} & 3194 & 6033 & 4.0k & 5514  & 
    2641.6 s & 1488.8 s & 2609.5 s & 474.7 s &
    5.6$\times$ & 3.1$\times$ & 5.5$\times$ \\
    
    \texttt{JPEG} & 3826 & 56785 & 2.0k & 5000  & 
    7214.8 s & 1450.0 s & 3945.7 s & 592.0 s &
    11.3$\times$ & 2.3$\times$ & 6.7$\times$ \\

     \texttt{Conv2D} & 11129 & 22258 & 4.0k & 5000  & 
    3271.2 s & 2050.0 s & 4104.2 s & 450.3 s & 
    7.3$\times$ & 4.6$\times$ & 9.1$\times$ \\
    
    % \texttt{StreamComp} & 31489 & 0 & 1.0k & 5514  & 
    % 0 s & 0 & 0 s & 0 s &
    % 3.2$\times$ & 3.2$\times$ & 3.8$\times$ \\
    \midrule
    Average & \multicolumn{1}{c}{-} & \multicolumn{1}{c}{-} & \multicolumn{1}{c}{-} & \multicolumn{1}{c}{-}& \multicolumn{4}{|c|}{-} & 7.0$\times$ & 11.4$\times$ & 6.1$\times$ \\
    \bottomrule
    \bottomrule
    \multicolumn{12}{l}{\#VVPs: The total number of lines of code generated by compiling all behavioral-level code into VVP assembly code, shown in Figure \ref{iverilog_example}. }
  \end{tabular}
  }
  \label{performance_cmp}
\end{table*}

Fig. \ref{sync_hidden} illustrates an example. Dashed lines represent variable synchronization, solid lines indicate dependency relationships, red boxes denote global synchronization, and green boxes mark the local synchronization of node \textit{e}. 
Suppose node \textit{e} contains a non-blocking assignment that updates the value of node \textit{a}. 
Following our approach, we first add a local synchronization node~(as shown by the green box) for node \textit{e} and then identify tasks dependent on the signal to be synchronized. Since node \textit{d} depends on the value of node \textit{a}, we add a dependency edge from the node \textit{d} to the local synchronization node. This dependency edge guarantees that the signal to be synchronized has been fully utilized before the local synchronization task is executed.
% , ensuring the safe and correct execution of the synchronization process. 
Furthermore, assuming uniform execution times across all nodes in the task graph, the local synchronization approach allows the synchronization of node \textit{a} to occur in parallel with the execution of node \textit{f}, effectively hiding the synchronization overhead associated with node \textit{a}.  

% \begin{algorithm}
% \caption{Execution Flow Redundancy Elimination} \label{alg_execution}
% \KwIn{ Behavior\_code: A fault-free behavioral code; fault\_id: Specified the faulty behavioral code}
% \KwOut{ Whether the faulty behavioral code is redundant }
% \end{algorithm} 

% \subsection{Task Execution }

\section{Evaluation}
\subsection{\textbf{Experimental Settings}}
We implement RIROS in C++ using the Taskflow parallelization framework\cite{taskflow} and execute it on a host equipped with an Intel(R) Core(TM) i7 processor, which features 16 cores operating at 3.80GHz.
% , and a RAM capacity of 32GB. 
We use gcc-11 to compile the simulator and enable the -O2 optimization option. 
A threshold of 0.0001 is used to identify high-load nodes, as it achieves optimal performance in our experiments.

% 据我们所知，RIROS是第一个并行RTL故障仿真框架。为了评估我们的性能，我们实现了三个不同版本的RIROS，分别是RIROS-static, RIROS-dynamic and chamelemon，其中RIROS-static采用静态划分方法，RIROS-dynamic采用动态划分方法，RIROS采用动态划分和全局同步优化。为了方便起见，我们接下适用RIROS\textbf{-}\textbf{-}, RIROS-, RIROS分别来表示。
% To the best of our knowledge, RIROS is the first parallel RTL fault simulation framework. 
% 为了评估RIROS的性能，我们与ERASER和VCS两个仿真器做对比。其中，ERASER是一个RTL故障仿真器，VCS则是通过使用force/release命令模拟故障仿真。例如，当需要模拟一个Wire上的SA0故障时，我们在testbench中的initial代码块中插入force tb.dut.xxx = 0指令强制某根wire的值为0. 
To evaluate the performance of RIROS, we compare it with two simulators: ERASER\cite{ERASER}, a dedicated single-core RTL fault simulator, and VCS\cite{vcs}, which emulates fault simulation using \textit{force/release} commands. For instance, to simulate a SA0 fault on a wire \textit{n1}, we insert a \textit{force tb.dut.n1 = 0} statement into the initial block of the testbench, which forces the target wire to hold a constant value of 0.
To ensure a fair comparison with VCS, all faults are retained without any being dropped in RIROS and ERASER.
% For simplicity, these variants are referred to as RIROS\textbf{-}\textbf{-}, RIROS-, and RIROS, respectively, in the following discussion.

\textbf{Benchmark.} 
The benchmarks, as shown in Table \ref{performance_cmp}, used in this paper cover video controllers\cite{OpenCores}, RISCV CPUs\cite{ucb_bar, picorv32}, and accelerators\cite{conv_1, tensorlib}. 
% For the SDRAM controller design, due to limitations in our current simulator's support for inout ports, we artificially partitioned the inout port into distinct input and output ports. 
% In our experiments, the stimuli used are either sourced from test benches provided by the design developers or written by us based on the function of benchmarks if test benches are not provided. 
% The simulation cycles for each design are also shown in Table \ref{benchmark}. 
We generate stuck-at faults for wires and regs in the designs.
% 需要注意的是，根据该我们的统一故障注入方法，as shown in Section xxx，该框架只需要重写FaultyVal函数即可支持其他任意故障模型，如SET或SEU瞬态故障模型。
Notably, as described in Section \ref{fault_injection}, our unified fault injection methodology enables RIROS to support a wide range of fault models, such as transient faults\cite{transient_fault}, by simply overriding the \textit{FaultyVal} function.

% \begin{table}[h]
% \centering
% \setlength{\abovecaptionskip}{0cm}
% \caption{Benchmark Information}

% \label{benchmark}
% \resizebox{0.9\linewidth}{!}{
% \begin{tabular}{|c|c|c|c|}
% \hline
% \textbf{Benchmark} &
% \textbf{\#Stimulus} &
% \textbf{\#Cells} &
% \textbf{\#Faults}
% \\
% \hline
% \hline
%  SHA256 &  &  &    \\
% \hline
%  JPEG &  &  &   \\
% \hline
% Sodor Core & 3k & 16943 & 1252  \\
% \hline
% RISCV Mini & 6k & 9087 &  526  \\
% \hline
% PicoRV32 & 4k  & 17488 & 1040   \\
% \hline
% Conv\_acc & 4k & 39812 & 1032  \\
% \hline
% GEMM & 4k &   &    \\
% \hline
% % Average & \multicolumn{3}{c|}{-} & \multicolumn{2}{c}{Consistency} \\
% % \hline
% \multicolumn{4}{l}{\textbf{\#Cells}: Number of cells reported by the Yosys tool\cite{Yosys}. } \\
% % \multicolumn{6}{l}{\textbf{$\ast$ SHA256\_HV}: the handwritten Verilog of SHA256. } \\
% % \multicolumn{6}{l}{\textbf{$\star$ SHA256\_C2V}: the Verilog generated by Chisel of SHA256. }
% \end{tabular}
% }
% % \caption*{
% %  \#Cells: Number of cells reported by the Yosys tool\cite{Yosys}.
% %  The SHA256\_HV is implemented by handwritten Verilog. The verilog of SHA256\_C2V is generated by Chisel. 
% % }
% \end{table}

\subsection{\textbf{Overall Performance}}\label{overall_performance}
% RIROS-static RIROS-dynamic Chaemelon 三个仿真器

% 我们对上述所有 benchmark 分别使用 ERASER、VCS 和 RIROS 三种仿真器执行RTL故障仿真并进行了性能对比。表 xxx 展示了ERASER，VCS和RIROS单核，RIROS 16核在不同 benchmark 的执行时间。例如，对于PicoRV32设计，ERASER需要使用449.9s完成6898个故障的故障仿真，VCS需要使用1448.6s，RIROS单核需要使用301.5s，而RIROS 16核只需要使用56.4s。
% 在大部分电路上, ERASER与RIROS单核性能相等。而对于JPEG设计，RIROS获得1.8倍的性能加速，可能的原因是RIROS消除了事件队列的管理，避免事件的入队和出队开销。
% 另外，在一些电路上，RIROS单核的性能也优于VCS。比如，对于Conv Acc设计，相比于VCS，RIROS单核能够获得3.8倍的性能加速。
We performed RTL fault simulation on all the aforementioned benchmarks using three simulators: ERASER, VCS, and RIROS, and compared their performance. Table \ref{performance_cmp} presents the execution times for ERASER, VCS, RIROS (single-core), and RIROS (16-core) across various benchmarks.
For example, when simulating the PicoRV32 design with injecting 6,898 faults, ERASER takes 449.9 s to finish, VCS takes 1,448.6 s, RIROS (single-core) takes only 301.5 s, and RIROS (16-core) reduces the time to just 56.4 s. 
In most cases, RIROS (single-core) performs similarly to ERASER. However, for the JPEG design, RIROS achieves a 1.8$\times$ performance speedup, likely due to eliminating the overhead of event queue management, such as enqueuing and dequeuing costs.
Besides, in some circuits, RIROS (single-core) outperforms VCS. For instance, in the Conv Acc design, RIROS (single-core) achieves a 3.8$\times$ speedup over VCS.

% 在多核加速下，相比于ERASER，RIROS 16核在仿真JPEG设计时获得11.3倍的性能加速。相比于VCS，RIROS 16核在仿真PicoRV32时获得25.7倍的性能加速。相比于RIROS 单核，RIROS 16核在仿真Conv2D时获得9.1倍的性能加速。
% 平均而言，相比于ERASER，VCS，RIROS单核，RIROS 16核分别获得7.0倍，11.4倍，6.0倍的性能加速。
% 这充分体现出了RIROS框架的高效性。
With multi-core acceleration, RIROS (16-core) achieves an 11.3$\times$ performance speedup in simulating the JPEG design compared to ERASER. When compared to VCS, RIROS (16-core) delivers a 25.7$\times$ speedup in simulating PicoRV32. Additionally, RIROS (16-core) achieves a 9.1$\times$ speedup over RIROS (single-core) in simulating Conv2D.
On average, RIROS (16-core) delivers performance speedups of 7.0$\times$, 11.4$\times$, and 6.1$\times$ compared to ERASER, VCS, and RIROS (single-core), respectively. These results clearly highlight the high efficiency of the RIROS framework.

\begin{figure}[htbp]
    \centering
    % \setlength{\abovecaptionskip}{-0.00cm}
    % \setlength{\belowcaptionskip}{-0.00cm}
    % \includesvg[inkscapelatex=false, width=0.75\linewidth]{figure/overall_speedup2.svg}
    \includegraphics[width=0.75\linewidth]{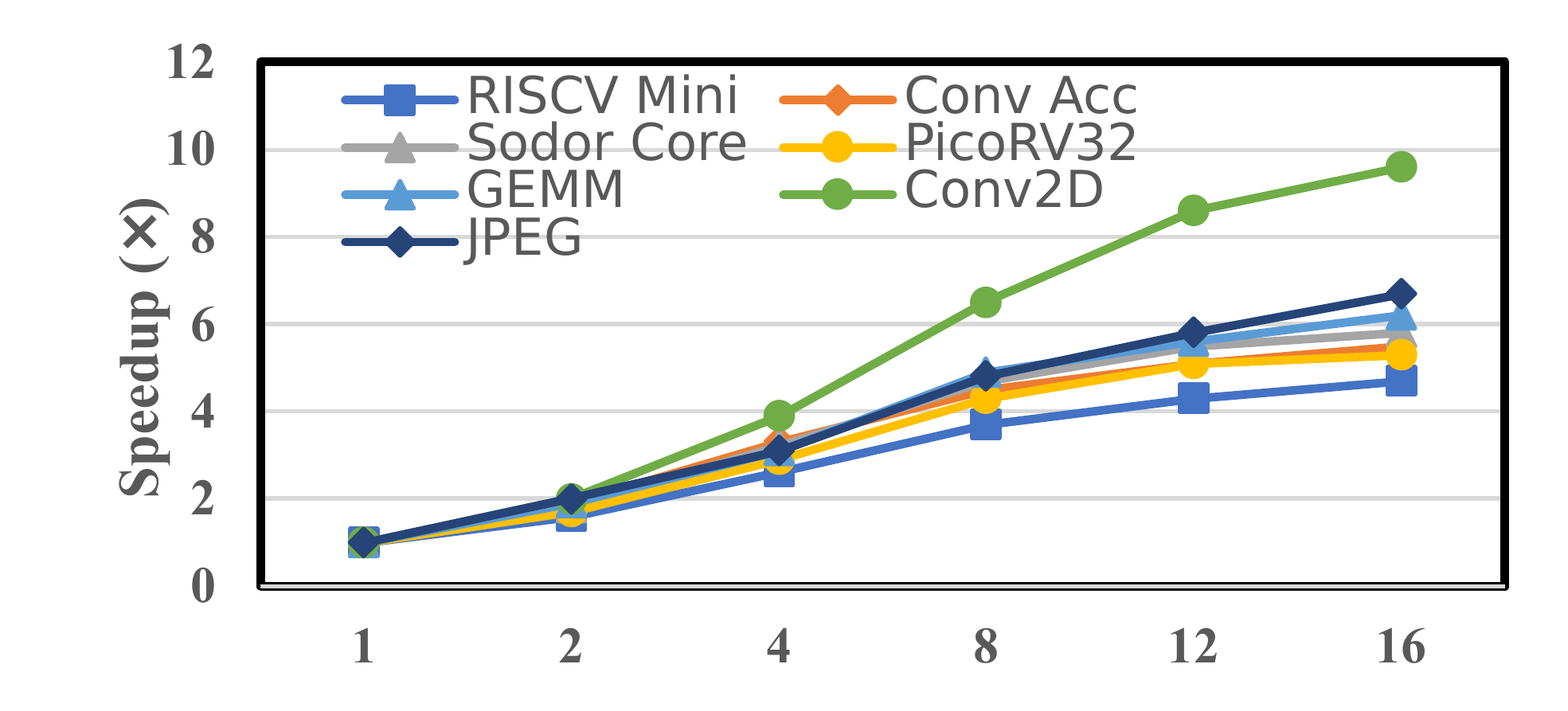}
    \caption{
       The speed-ups of all designs across different cores.
    }
    \label{speedup_diff_cores}
\end{figure}

Fig. \ref{speedup_diff_cores} presents the speedup for all benchmarks under different cores to further investigate the scalability of the RIROS parallel framework. 
% 可以看到，随着core数量的增加，RIROS的性能也在不断提升，这表明RIROS具有很好的可扩展性。这得益于该论文提出的两个优化策略，这将在消融实验中得到证实。
% 更重要的是，电路的规模越大，RIROS的加速比也越大，如Conv2D电路在16核下能够获得9.1的加速比。这表明RIROS在大电路上的扩展性更好。
As shown, RIROS demonstrates strong scalability, as its performance consistently improves with an increasing number of cores. This phenomenon is attributed to the two optimization strategies proposed in the paper, which will be confirmed in the ablation experiments presented later.
More importantly, the speedup becomes more significant with larger designs. For instance, the Conv2D design achieves a 9.1$\times$ speedup when running on 16 cores, highlighting RIROS's superior scalability on large-scale designs.

\subsection{\textbf{Ablation Study}}

To evaluate the effectiveness of the different optimization strategies proposed in this paper, we implemented two variants: RIROS\textbf{-}\textbf{-} and RIROS\textbf{-}. The details are as follows: \textbf{RIROS\textbf{-}\textbf{-}} implements structural-level parallelism; \textbf{RIROS\textbf{-}} implements structural-level parallelism and fault-level parallelism for overloaded nodes. 
% \begin{itemize}
%     \item \textbf{RIROS\textbf{-}\textbf{-}} implements structural-level parallelism.
%     \item \textbf{RIROS\textbf{-}} implements structural-level parallelism and fault-level parallelism for overloaded nodes. 
% \end{itemize}

% \subsubsection{\textbf{Task-based simulation}}
% % 

\begin{figure}[htbp]
    \centering
    % \setlength{\abovecaptionskip}{-0.00cm}
    % \setlength{\belowcaptionskip}{-0.00cm}
    % \includesvg[inkscapelatex=false, width=0.8\linewidth]{figure/abs_speedup2.svg}
    \includegraphics[width=0.8\linewidth]{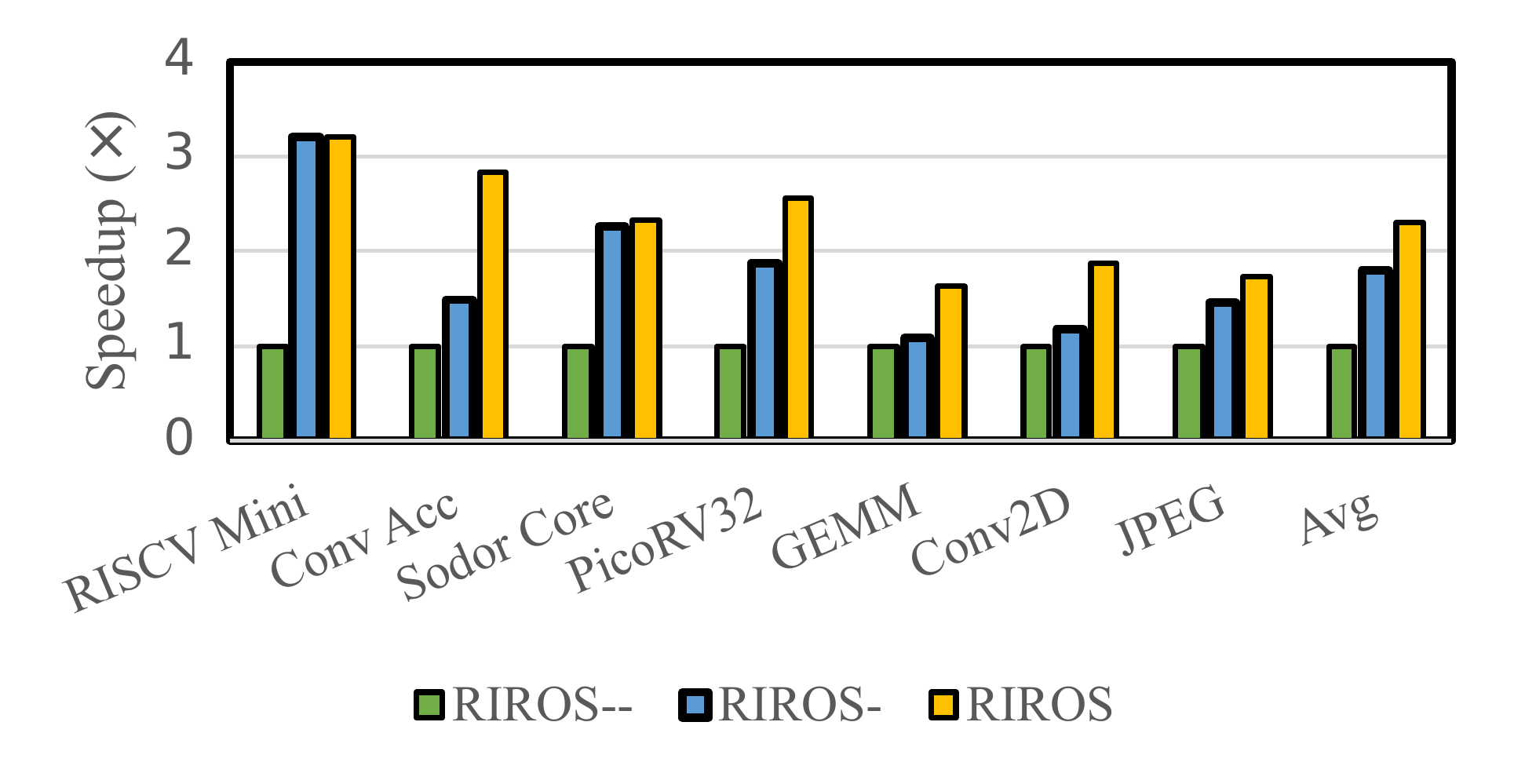}
    \caption{ 
        16-core runtime of three versions, the RIROS\textbf{-}\textbf{-} is used as the baseline.
    }
    \label{abs_speedup}
\end{figure}

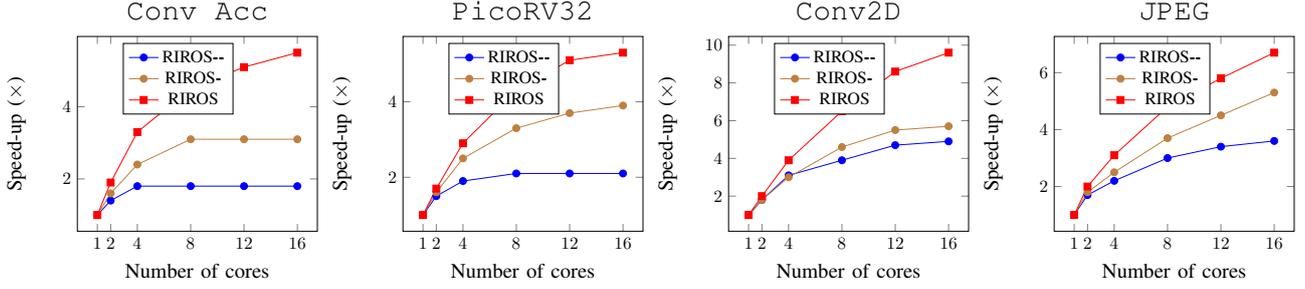
\begin{figure*}[htbp]
  \centering
  \pgfplotsset{
    title style={font=\LARGE},
    label style={font=\large},
  }
  % \begin{tikzpicture}[scale=0.65]
  %   \begin{axis}[
  %     title=\texttt{RISCV Mini},
  %     ylabel=Speed-up ($\times$),
  %     xlabel=Number of cores,
  %     width=0.35\linewidth,
	 %  legend style={at={(0.97,0.3)},anchor= south east},
  %     xtick={1, 2, 4, 8,  12, 16},
  %     cycle list={
  %           {blue, mark=*},        
  %           {brown, mark=*}, 
  %           {red, mark=square*},   
  %       }
  %     ]

  %     \addplot table {
  %       1  1.0
  %       2  1.3
  %       4  1.5
  %       8  1.5
  %       12 1.5
  %       16  1.5
  %     };
  %     \addlegendentry{RIROS\textbf{-}\textbf{-}}
      
  %     \addplot table {
  %       1  1.0
  %       2  1.6
  %       4  2.6
  %       8  3.8
  %       12  4.3
  %       16  4.7
  %     };
  %     \addlegendentry{RIROS\textbf{-}}

  %     \addplot table {
  %       1  1.0
  %       2  1.6
  %       4  2.6
  %       8  3.7
  %       12  4.3
  %       16  4.7
  %     };
  %     \addlegendentry{RIROS}

  %     % \legend{RIROS-, RIROS}
  %   \end{axis}
  % \end{tikzpicture}
  \begin{tikzpicture}[scale=0.65]
    \begin{axis}[
      title=\texttt{Conv Acc},
      ylabel=Speed-up ($\times$),
      xlabel=Number of cores,
      width=0.35\linewidth,
	  legend style={at={(0.65,0.6)},anchor= south east},
      xtick={1, 2, 4, 8,  12, 16},
      cycle list={
            {blue, mark=*},        
            {brown, mark=*}, 
            {red, mark=square*},   
        }
      ]

      \addplot table {
        1  1.0
        2  1.4
        4  1.8
        8  1.8
        12 1.8
        16  1.8
      };
      \addlegendentry{RIROS\textbf{-}\textbf{-}}
      
      \addplot table {
        1  1.0
        2  1.6
        4  2.4
        8  3.1
        12  3.1
        16  3.1
      };
      \addlegendentry{RIROS\textbf{-}}

      \addplot table {
        1  1.0
        2  1.9
        4  3.3
        8  4.5
        12  5.1
        16  5.5
      };
      \addlegendentry{RIROS}

      % \legend{RIROS-, RIROS}
    \end{axis}
  \end{tikzpicture}
  \begin{tikzpicture}[scale=0.65]
    \begin{axis}[
      title=\texttt{PicoRV32},
      ylabel=Speed-up ($\times$),
      xlabel=Number of cores,
      width=0.35\linewidth,
	  legend style={at={(0.65,0.6)},anchor= south east},
      xtick={1, 2, 4, 8,  12, 16},
      cycle list={
            {blue, mark=*},        
            {brown, mark=*}, 
            {red, mark=square*},   
        }
      ]

      \addplot table {
        1  1.0
        2  1.5
        4  1.9
        8  2.1
        12  2.1
        16  2.1
      };
      \addlegendentry{RIROS\textbf{-}\textbf{-}}
      
      \addplot table {
        1  1.0
        2  1.6
        4  2.5
        8  3.3
        12  3.7
        16  3.9
      };
      \addlegendentry{RIROS\textbf{-}}

      \addplot table {
        1  1.0
        2  1.7
        4  2.9
        8  4.3
        12  5.1
        16  5.3
      };
      \addlegendentry{RIROS}

      % \legend{RIROS-, RIROS}
    \end{axis}
  \end{tikzpicture}
  \begin{tikzpicture}[scale=0.65]
    \begin{axis}[
      title=\texttt{Conv2D},
      ylabel=Speed-up ($\times$),
      xlabel=Number of cores,
      width=0.35\linewidth,
	  legend style={at={(0.65,0.6)},anchor= south east},
      xtick={1, 2, 4, 8,  12, 16},
      cycle list={
            {blue, mark=*},        
            {brown, mark=*}, 
            {red, mark=square*},   
        }
      ]

      \addplot table {
        1  1.0
        2  1.8
        4  3.1
        8  3.9
        12  4.7
        16  4.9
      };
      \addlegendentry{RIROS\textbf{-}\textbf{-}}
      
      \addplot table {
        1  1.0
        2  1.8
        4  3.0
        8  4.6
        12  5.5
        16  5.7
      };
      \addlegendentry{RIROS\textbf{-}}

      \addplot table {
        1  1.0
        2  2
        4  3.9
        8  6.5
        12  8.6
        16  9.6
      };
      \addlegendentry{RIROS}

      % \legend{RIROS-, RIROS}
    \end{axis}
  \end{tikzpicture}
  \begin{tikzpicture}[scale=0.65]
    \begin{axis}[
      title=\texttt{JPEG},
      ylabel=Speed-up ($\times$),
      xlabel=Number of cores,
      width=0.35\linewidth,
	  legend style={at={(0.65,0.6)},anchor= south east},
      xtick={1, 2, 4, 8, 12, 16},
      cycle list={
            {blue, mark=*},        
            {brown, mark=*}, 
            {red, mark=square*},   
        }
      ]

      \addplot table {
        1  1.0
        2  1.7
        4  2.2
        8  3.0
        12  3.4
        16  3.6
      };
      \addlegendentry{RIROS\textbf{-}\textbf{-}}
      
      \addplot table {
        1  1.0
        2  1.8
        4  2.5
        8  3.7
        12 4.5
        16  5.3
      };
      \addlegendentry{RIROS\textbf{-}}

      \addplot table {
        1  1.0
        2  2.0
        4  3.1
        8  4.8
        12 5.8
        16  6.7
      };
      \addlegendentry{RIROS}

      % \legend{RIROS-, RIROS}
    \end{axis}
  \end{tikzpicture}

  \caption{The speedup of RIROS\textbf{-}\textbf{-}, RIROS\textbf{-} and RIROS across different cores.} 
  \label{speedup}
\end{figure*}
\begin{figure*}[htbp]
    \centering
    % \setlength{\abovecaptionskip}{-0.00cm}
    % \setlength{\belowcaptionskip}{-0.00cm}
    % \includesvg[inkscapelatex=false, width=\linewidth]{figure/CPU_utilization_8_2.svg}
    \includegraphics[width=0.9\linewidth]{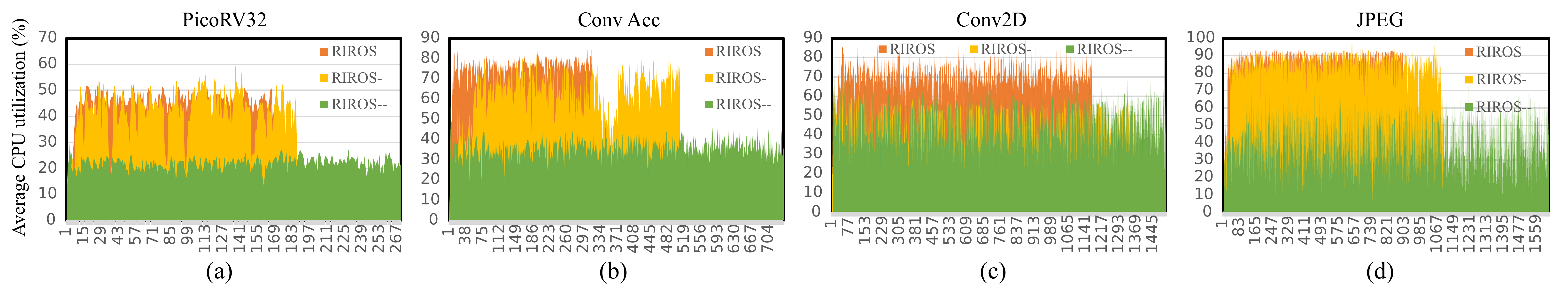}
    \caption{ 
    Profiling the average CPU utilization and the time sampling window is 1s. The x-axis represents the sampling points. Note that the runtime includes profiling overhead.
    % 注意：这里的runtime与Table xx存在区别，是因为profile会产生额外的开销。
    % Note that the runtime differs from Table \ref{performance_cmp} due to profiling overhead.
    % Note that the runtime includes profiling overhead.
    }
    \label{cpu_utilization}
\end{figure*}
\begin{figure}[htbp]
    \centering
    % \setlength{\abovecaptionskip}{-0.00cm}
    % \setlength{\belowcaptionskip}{-0.00cm}
    % \includesvg[inkscapelatex=false, width=\linewidth]{figure/picorv_balance.svg}
    \includegraphics[ width=\linewidth]{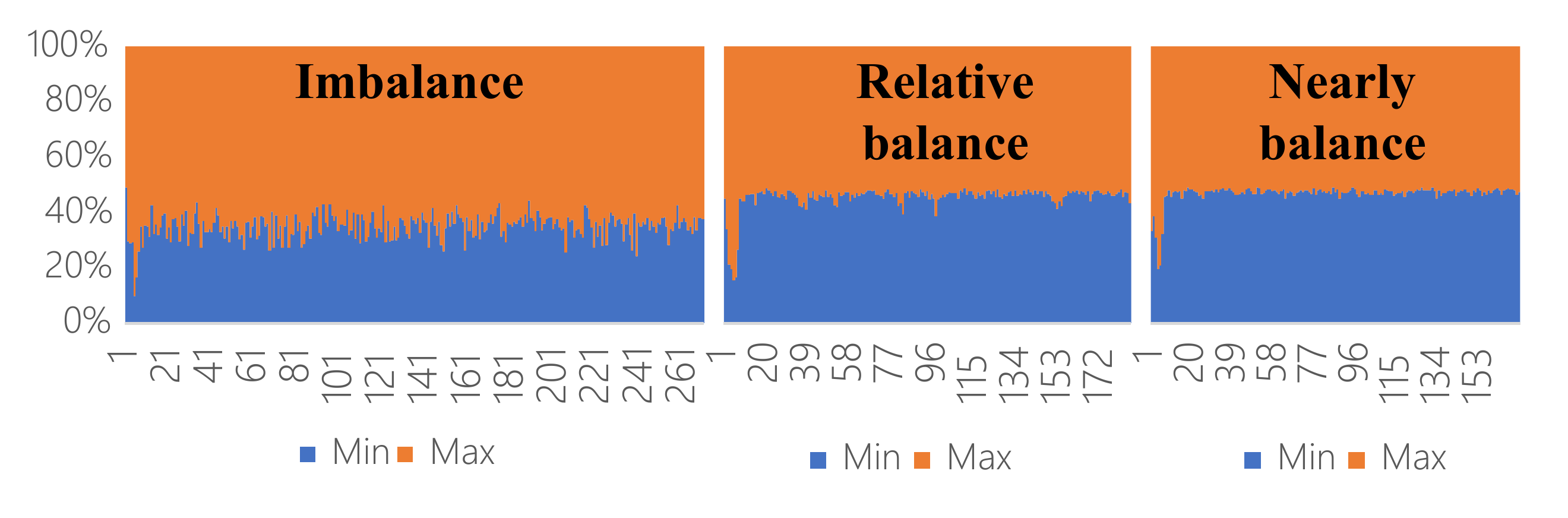}
    \caption{ 
         The Maximum and Minimum CPU utilization of PicoRV32 under 8 cores. 
        (a) RIROS\textbf{-}\textbf{-}; 
        (b) RIROS\textbf{-};
        (c) RIROS.
    }
    \label{balance_cpu_utilization}
\end{figure}
% 实验结论是什么
% RISCV Mini 换成 Conv Acc
\subsubsection{\textbf{Effectiveness of Multi-Dimensional Parallelism}}
% 16核下的绝对性能对比

% 我们通过对比RIROS--,RIROS-和RIROS在不同设计的运行时间来验证我们方法的有效性。结果在xxx图中展示（RIROS--的运行时间作为baseline），可以看到，RIROS-的性能总是比RIROS-更好，尤其是对于RISCV Mini这个设计，获得3.2倍的性能提升。这表明了增加故障维度的并行能够带来很好的性能收益。然而，我们发现，对于GEMM和Conv2D两个设计来说，其增加故障维度的并行带来的收益相对有限。这是因为其计算计算阶段的其不同核之间的负载相对均匀，最高核利用率与最低核利用率差距分别为xxx和xxx。
% 此外，对比RIROS-和RIROS可以发现，加速效果能够进一步提升。如对于PicoRV32设计，加速比从1.9增加到2.7。这表明增加操作维度的并行也能够性能收益。
% 总体而言，相比于结构级并行，同时增加故障级并行和操作级并行能够带来2.3倍的性能提升。
We evaluate the effectiveness of our approach by comparing the runtime performance of RIROS\textbf{-}\textbf{-}, RIROS\textbf{-}, and RIROS across a range of benchmark designs, as shown in Figure \ref{abs_speedup}. Here, RIROS\textbf{-}\textbf{-} serves as the baseline.
Overall, RIROS\textbf{-}, which introduces fault-dimension parallelism on top of structural-level parallelism, consistently outperforms RIROS\textbf{-}\textbf{-}. The improvement is particularly notable in the RISCV Mini design, where we observe a 3.2$\times$ speedup. This demonstrates that exploiting intra-task parallelism among fault dimension (i.e., bad gate parallelism) can significantly reduce simulation time.
% However, the gains from fault-dimension parallelism are less pronounced for GEMM and Conv2D. This is primarily due to the balanced workload across cores during the compute phase in these designs. Specifically, the difference between the highest and lowest core utilization rates is only XXX and XXX, respectively, leaving limited idle time to exploit.
Further improvements are observed when comparing RIROS\textbf{-} with RIROS, which additionally introduces operation-level parallelism by interleaving computation and synchronization. For instance, in the PicoRV32 design, the speedup increases from 1.9$\times$ to 2.7$\times$, showing that fine-grained synchronization further improves performance.
In summary, by complementing structural-level parallelism with both fault-level and operation-level parallelism, our method achieves up to a 2.3$\times$ average speedup, demonstrating its effectiveness in improving RTL fault simulation performance across diverse workloads.

% 为了进一步对比不同并行维度的扩展性，我们对比了RIROS--,RIROS-和RIROS在不同core下的加速效果。图xxx展示了不同电路在不同core的加速比。可以看到，在所有电路上，多维度并行不仅获得最高的加速比，扩展性也更好。比如对于Conv2D电路，结构并行，在16核下，只能获得3.6倍的加速且性能已趋于稳定，而我们提出的多维并行能够能够获得8.6倍的加速且依旧存在上升空间。
To further assess the scalability of different parallel dimensions, we compared the acceleration performance of RIROS\textbf{-}\textbf{-}, RIROS\textbf{-}, and RIROS across various core counts. Figure \ref{speedup} illustrates the speedup of different designs for different core configurations. The results clearly show that RIROS not only delivers the highest speedup but also exhibits superior scalability across all circuits. For example, in the case of the Conv2D circuit, RIROS\textbf{-}\textbf{-} achieves a speedup of only 4.7$\times$ at 12 cores, with performance reaching a plateau. In contrast, RIROS achieves a 8.6$\times$ speedup, with potential for further gains.

\subsubsection{\textbf{CPU Utilization Comparison}}
% CPU利用率展示
%我们使用 Intel VTune 工具进一步分析xxx、xxx和xxx的平均CPU利用率。如图xxx展示了4个设计，在8核仿真，的平均CPU利用情况随时间的变化。
% 可以看到，RIROS--的整体CPU利用率很低。如对于PicoRV32电路，CPU利用率平均只有22\%，这说明结构级并行存在大量的空闲核。与RIROS--相比，RIROS-的CPU利用率得到有效改善。这是因为故障维度的并行能够用使用小任务去填充空闲的核，从而提升核的利用率。进一步，通过打破计算-全局同步边界，能够使用全局同步中的任务进一步填充核的空闲，从而进一步提高与RIROS-的CPU利用率。
We use the Intel VTune\cite{vtune} tool to further analyze the average CPU utilization of RIROS\textbf{-}\textbf{-}, RIROS\textbf{-}, and RIROS. Figure \ref{cpu_utilization} illustrates the time-varying CPU utilization of four different designs during 8-core simulations.
The results show that RIROS\textbf{-}\textbf{-} exhibits consistently low CPU utilization. For instance, in the case of the PicoRV32 circuit, the average CPU utilization is only 22\%, indicating a substantial number of idle cores when relying solely on structural-level parallelism. In comparison, RIROS\textbf{-} achieves significantly better utilization. This improvement stems from the introduction of fault-dimension parallelism, which enables small tasks to be distributed across idle cores, thereby increasing overall resource efficiency.
RIROS takes this a step further by breaking the boundary between the computation and global synchronization phases. By allowing synchronization-phase tasks to occupy idle cores, it further enhances CPU utilization beyond that of RIROS\textbf{-}. These results highlight the clear advantages of multi-dimensional parallelism in maximizing core utilization and improving simulation efficiency.

% 负载平衡评估
\subsubsection{\textbf{Load Balancing Evaluation}}
% 我们进一步研究了xxx，xxx和xxx在负载平衡方面的能力。如图xxx展示了xxx电路的最大和最小CPU利用率。可以看到，xxx的最大最小利用率相差xxx,而xxx只有xxx。这表明xxx在负载均衡方面具有更优表现，能够更有效地在多个核心之间均衡分配任务，从而提升整体并行效率。
We evaluated the load balancing of RIROS\textbf{-}\textbf{-}, RIROS\textbf{-}, and RIROS. As shown in Figure \ref{balance_cpu_utilization}, the gap between maximum and minimum CPU utilization for the PicoRV32 design gradually narrows. This indicates that RIROS achieves better load distribution across cores, leading to improved parallel efficiency.

\begin{figure}[htbp]
    \centering
    % \setlength{\abovecaptionskip}{-0.00cm}
    % \setlength{\belowcaptionskip}{-0.00cm}
    % \includesvg[inkscapelatex=false, width=\linewidth]{figure/cost3.svg}
    \includegraphics[width=\linewidth]{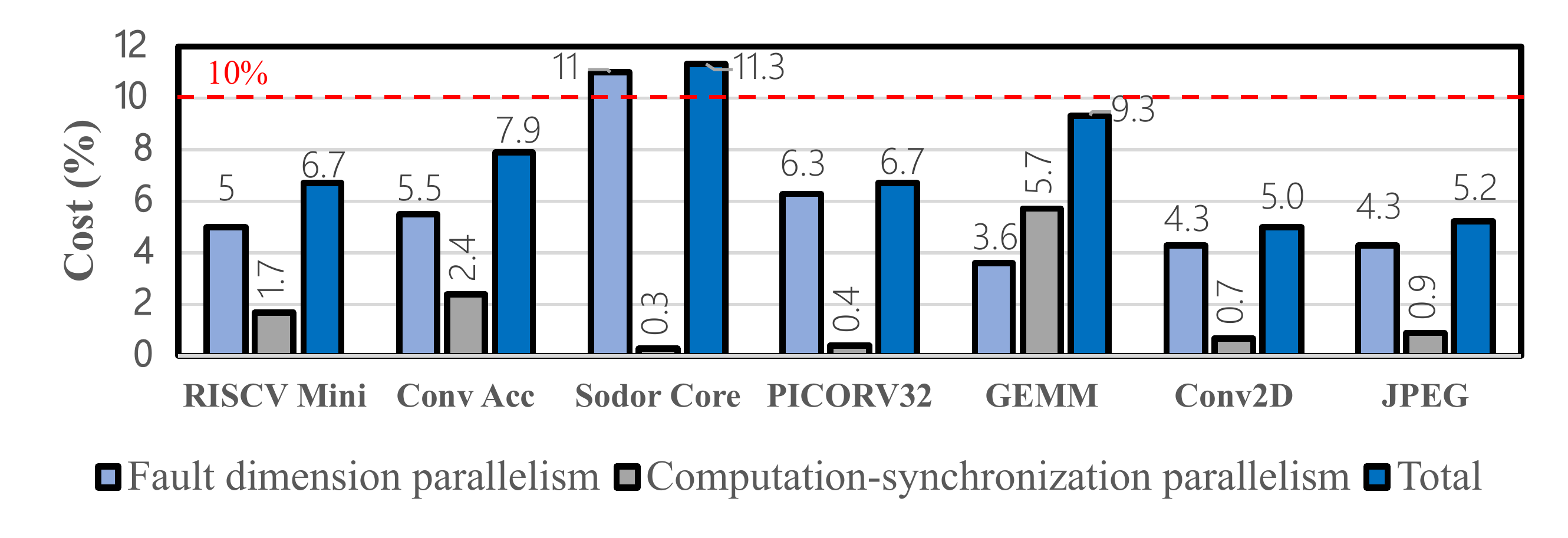}
    \caption{The cost introduced by the fault dimension parallelism and computation-global synchronization parallelism. 
    }
    \label{cost}
\end{figure}
\subsection{\textbf{Cost Analysis}}\label{profile_analysis}

As shown in Section~\ref{framework}, both fault-level parallelism and computation-global synchronization parallelism are achieved by adding task nodes and dependency edges to the task graph, introducing extra task scheduling overhead and dependency resolving overhead. 
% 在cost测试的时候，我们并不区分任务调度开销和依赖resolve开销，而是将其看成一个整体。
% 图xxx展示了动态任务重划分所引入的slave task的开销、局部同步优化所引入的local sync task的开销及整体的开销 for each benchmark on 16 cores。例如，对于JPEG设计，总体引入5.2%的开销，其中动态任务划分贡献4.3%，而局部同步优化贡献0.9.总体而言，在大部分设计上，该论文提出的优化策略的引入的开销都低于10%，体现了优化策略的轻量性。
% 
We do not distinguish between task scheduling overhead and dependency resolution overhead, as the two are not entirely separable. 
% Instead, they are considered as a unified cost. 
Figure \ref{cost} illustrates the overhead introduced by fault dimension parallelism, computation-global synchronization parallelism, and the total overhead across various benchmarks under 16 cores. 
For example, for the JPEG design, the total overhead is 5.2\%, with 4.3\% attributed to fault dimension parallelism and 0.9\% to computation-global synchronization parallelism. 
Overall, the overhead introduced by the proposed optimization strategies remains low, highlighting the efficiency and lightweight nature of our approaches.

\section{Conclusion}
In this paper, we implement a parallel RTL fault simulation framework, RIROS, which integrates structural-level and fault-level parallelism approaches to minimize bubbles
that arise from parallel core load imbalances. 
Additionally, we introduced a unified scheduling method to further fill bubbles. 
The experimental results show that RIROS achieves a 7.0$\times$ and 11.0$\times$ performance improvement compared to the state-of-the-art RTL fault simulation and a commercial tool.

\section*{Acknowledgment}
This paper is supported in part by the National Natural Science Foundation of China (NSFC) under grant No.~(92473203, 92373206), and in part by the Chinese Academy of Sciences under grant No. XDB0660100.
The corresponding authors are Jianan Mu and Huawei Li.

% \section*{Acknowledgment}

% \newpage

% \vspace{12pt}
% \color{red}
% IEEE conference templates contain guidance text for composing and formatting conference papers. Please ensure that all template text is removed from your conference paper prior to submission to the conference. Failure to remove the template text from your paper may result in your paper not being published.

\end{document}